\documentclass{article}
\usepackage[utf8]{inputenc}

\usepackage{amsthm,amsmath,bm,bbm}
\usepackage{amssymb,mathtools}
\usepackage{fullpage}
\usepackage[dvipsnames]{xcolor}
\usepackage[colorlinks=true, allcolors=violet]{hyperref}

\usepackage{multirow}
\usepackage{caption}
\usepackage{booktabs}

\usepackage{enumitem}

\usepackage{natbib}
\setcitestyle{authoryear, open={(},close={)}}
\usepackage{minitoc}
\makeatletter
\renewcommand{\paragraph}{%
  \@startsection{paragraph}{4}%
  {\z@}{.5ex \@plus 1ex \@minus .2ex}{-1em}%
  {\normalfont\normalsize\bfseries}%
}
\makeatother

\usepackage{tikz}
\usetikzlibrary{shapes,shadows,arrows,positioning,arrows.meta,decorations.pathreplacing,decorations.pathmorphing,decorations.shapes}
\usepackage{cancel}
\usepackage[colorinlistoftodos,textsize=scriptsize,textwidth=.8in,color=blue!20]{todonotes}

\newlength\tindent
\def\independent{\perp\!\!\!\perp}

\def\E{\text{E}}
\def\P{\text{P}}

\def\S{\mathcal{S}}

\colorlet{no}{cyan!200}
\colorlet{ok}{black!70}
\colorlet{rmod}{gray}
\colorlet{butx}{blue}
\colorlet{uhoh}{red}
\colorlet{half}{blue!50!red}

\allowdisplaybreaks

\title{In defense of MAR over latent ignorability (or latent MAR)\\for outcome missingness in studying principal causal effects:\\a causal graph view}

\author{Trang Quynh Nguyen}

\begin{document}

\maketitle

\begin{abstract}
    \noindent 
    This paper concerns outcome missingness in principal stratification analysis%, which targets principal causal effects
    . 
    % As the settings are similar, the findings here are also relevant to instrumental variable analysis that targets the local average treatment effect. 
    We revisit a common assumption known as \textit{latent ignorability} or \textit{latent missing-at-random} (LMAR), often considered a relaxation of missing-at-random (MAR). LMAR posits that the outcome is independent of its missingness if one conditions on principal stratum (which is partially unobservable) in addition to observed variables. The literature has focused on methods assuming LMAR (usually supplemented with a more specific assumption about the missingness), without considering the theoretical plausibility and necessity of LMAR. In this paper, we devise a way to represent principal stratum in causal graphs, and use causal graphs to examine this assumption. We find that LMAR is harder to satisfy than MAR, and for the purpose of breaking the dependence between the outcome and its missingness, no benefit is gained from conditioning on principal stratum on top of conditioning on observed variables. This finding has an important implication: MAR should be preferred over LMAR. This is convenient because MAR is easier to handle and (unlike LMAR) if MAR is assumed no additional assumption is needed.  We thus turn to focus on the plausibility of MAR and its implications, with a view to facilitate appropriate use of this assumption.
    % To facilitate appropriate use of MAR, we then consider this assumption's plausibility and implications. 
    We clarify conditions on the causal structure and on auxiliary variables (if available) that need to hold for MAR to hold, and we use MAR to recover effect identification under two dominant identification assumptions (exclusion restriction and principal ignorability). We briefly comment on cases where MAR does not hold. 
    In terms of broader connections, most of the MAR findings are also relevant to classic instrumental variable analysis that targets the local average treatment effect; and the LMAR finding suggests general caution with assumptions that condition on principal stratum.

    ~

\end{abstract}

\doparttoc % Tell to minitoc to generate a toc for the parts
\faketableofcontents % Run a fake tableofcontents command for the partocs
\part{} % Start the document part
% \parttoc % Insert the document TOC
% \tableofcontents

\vspace{-3em}

\section{Introduction}\label{sec:intro}

Principal stratification \citep{frangakis2002PrincipalStratificationCausal} is an important framework for handling noncompliance and post-treatment events in the study of treatment effects. With noncompliance, treatment received may differ from treatment assigned, reducing interest in the standard average treatment effect (ATE) estimand. The framework instead targets effects within \textit{principal strata}, i.e., subgroups of study participants defined based on the combination of their potential values (had treatment been assigned and had control been assigned) of treatment received. These effects are called \textit{principal causal effects}. 
There are different approaches to identifying these effects, which supplement standard causal inference assumptions with different assumptions to handle the fact that principal stratum is only partially observable. One approach treats the assigned treatment as an \textit{instrumental variable}, assuming that it affects the outcome only through treatment received (\textit{exclusion restriction} on the outcome) \citep{Angrist1995,angrist1996IdentificationCausalEffects}. Another approach relies on \textit{principal ignorability}, the assumption that conditional on a set of covariates principal stratum is independent of the outcome in certain mixtures of principal strata \citep{jo2009UsePropensityScores,stuart2015AssessingSensitivityMethods,feller2017PrincipalScoreMethods,ding2017PrincipalStratificationAnalysis}. Under such assumptions, the principal causal effects are identified -- in the absence of missing data.

This paper attends to the issue of outcome missingness, which is common \citep{wood2004AreMissingOutcome} and complicates effect identification. Specifically, we revisit a missingness assumption called latent ignorability \citep{frangakis1999AddressingComplicationsIntentiontotreat} or latent missing-at-random \citep{peng2004ExtendedGeneralLocation,nguyen2024IdentificationComplierNoncomplier}. We will use the latter label with its abbreviation LMAR. This assumption was formally proposed in \cite{frangakis1999AddressingComplicationsIntentiontotreat}, and independently appeared in a model in \cite{baker1998AnalysisSurvivalData}. It has since been used by many authors \citep{baker2000AnalyzingRandomizedCancer,yau2001InferenceComplierAverageCausal,barnard2003PrincipalStratificationApproacha,peng2004ExtendedGeneralLocation,mealli2004AnalyzingRandomizedTrial,frangakis2004MethodologyEvaluatingPartially,dunn2005EstimatingTreatmentEffects,omalley2005LikelihoodMethodsTreatment,zhou2006ITTAnalysisRandomized,lui2008SampleSizeDetermination,taylor2009MultipleImputationMethods,chen2009IdentifiabilityEstimationCausal,jin2010ModifiedGeneralLocation,rubin2010DealingNoncomplianceMissinga,jo2010HandlingMissingData,lui2010NotesOddsRatio,sobel2012ComplianceMixtureModellinga,mealli2012RefreshingAccountPrincipal,elliott2013AccommodatingMissingnessWhena,chen2015SemiparametricInferenceComplier,baker2016LatentClassInstrumental,diazordaz2019LocalAverageTreatment,nguyen2024IdentificationComplierNoncomplier}, ourselves included (the last citation), with a wide range applications including cancer screening, sepsis prevention, needle exchange, school vouchers, job training, elderly volunteering, and others. There are different versions, but the gist of the assumption is that the missingness is independent of the outcome conditional on principal stratum in addition to observed variables. The intuition is that the principal strata are different types of people, so it is reasonable to allow their missingness models to vary.

LMAR can be seen as a \textit{base assumption} about the missingness (it is generally not sufficient to recover effect identification), often combined with a more \textit{specific assumption} about the missingness mechanism. Most of the works cited above take the instrumental variable approach to effect identification.
To handle outcome missingness, following \cite{frangakis1999AddressingComplicationsIntentiontotreat}, LMAR is typically combined with an exclusion restriction on the missingness (treatment assignment affects missingness only through treatment received); the two exclusion restrictions (on outcome and on missingness) combined are called \textit{compound exclusion restriction}. \cite{mealli2004AnalyzingRandomizedTrial} and \cite{jo2010HandlingMissingData} give the alternative to combine LMAR with the \textit{stable complier response} assumption (a mirror image of exclusion restriction on missingness).
\cite{nguyen2024IdentificationComplierNoncomplier} generalize the use of LMAR, accommodating different effect identification approaches (instrumental variable, principal ignorability and deviations from principal ignorability) and expanding the range of specific missingness assumptions. They also note an issue with the exclusion restriction on missingness and stable complier response assumptions, and propose a mitigation.

Despite this strong history, the plausibility and necessity of LMAR have not been rigorously examined. In this paper, we use causal graphs to explore LMAR versus other missingness types, with a non-survival outcome. This is inspired by recent works that use causal graphs to examine existing methods \citep{ding2015AdjustNotAdjust,steiner2016MechanicsOmittedVariable,ding2017InstrumentalVariablesBias,kim2019CausalStructureSuppressor,kim2021CausalGraphicalViews,kim2021GainScoresRevisited}. We find that where the outcome is LMAR, it is also missing at random (MAR); and for the purpose of breaking dependence between the outcome and its missingness, there is no benefit to be gained from conditioning on principal stratum in addition to treatment received. This suggests that one can let go of LMAR and all its companion specific missingness assumptions, and instead embrace MAR. If MAR is deemed unlikely, standard missing-not-at-random (MNAR) methods should be used rather than adopting LMAR.
With this insight, we explore MAR carefully, clarifying conditions for MAR to hold and how auxiliary variables can help, and using MAR to recover effect identification. 
Additionally, we discuss what our findings mean for conditional independence assumptions that involve the principal stratum variable.

The paper is organized as follows. Section~\ref{sec:prelim} introduces the setting (including estimand and identification assumptions) and defines the missingness types LMAR, MAR and MNAR. Section~\ref{sec:toolbox} gathers a graphical toolbox for the job at hand, including a causal graph that represents principal stratification and conditional graphs that zoom into the model at specific values of observed variables. Section~\ref{sec:lmar} uses these graphs to examine missingness types, with focus on when LMAR holds. Section~\ref{sec:mar} takes a deep dive into MAR. Section~\ref{sec:recovery} uses MAR to recover effect identification under the instrumental variable approach and the principal ignorability approach. Section~\ref{sec:conclusion} provides a discussion.

\section{Preliminaries}
\label{sec:prelim}

\subsection{Setting and principal causal effects}\label{subsec:setting}

Let $Z$ denote treatment assigned, $S$ denote treatment received (or more generally, a post-treatment variable of interest), $Y$ denote the outcome, and $X$ denote baseline (i.e., pre-treatment-assignment) covariates. For simplicity of presentation, we take $Z$ and $S$ to be binary variables.

Adopting the potential outcomes framework \citep{splawa-neyman1990ApplicationProbabilityTheory,rubin1974EstimatingCausalEffects}, for $z=1,0$, let $Y(z)$ and $S(z)$ denote the potential values of $Y$ and $S$ under assignment to treatment $z$. 
Principal strata (indicated by variable $C$) are defined based on the combination of $S(1)$ and $S(0)$. We consider the two-sided noncompliance setting where $S(1)$ and $S(0)$ can both be either 0 or 1 ($S(0)$ can be 1 because people can access the treatment elsewhere), with four principal strata,
\begin{align*}
    C=\begin{cases}
        \text{always-taker} & \text{if }S(1)=S(0)=1
        \\
        \text{never-taker} & \text{if }S(1)=S(0)=0
        \\
        \text{complier} & \text{if }S(1)=1,S(0)=0
        \\
        \text{defier} & \text{if }S(1)=0,S(0)=1
    \end{cases};
\end{align*}
and the one-sided noncompliance setting where $S(0)=0$, with two principal strata,
\begin{align*}
    C=\begin{cases}
        \text{complier} & \text{if }S(1)=1
        \\
        \text{noncomplier} & \text{if }S(1)=0
    \end{cases}.
\end{align*}
The estimands are \textit{principal causal effects}, i.e., average causal effects of treatment assignment on the outcome within principal strata, formally $\E[Y(1)-Y(0)\mid C]$. 
Following the literature, we use AACE, NACE, CACE and DACE to refer to effects on \textit{a}lways-takers, \textit{n}ever-takers/\textit{n}oncompliers, \textit{c}ompliers and \textit{d}efiers specifically, and use PCEs to refer to principal causal effects generally.

\subsection{Effect identification under full data}
\label{subsec:principal-assumptions}

We assume that the PCEs would be identified in the absence of missing data. Throughout, we adopt the usual causal inference assumptions \textit{consistency} and \textit{treatment assignment ignorability} and \textit{positivity} (see A0-A2 below). If treatment assignment is randomized, a stronger condition than A1 holds, $Z\independent(S(1),S(0),Y(1),Y(0),X^*)$ where $X^*$ consists of everything before randomization, but here we consider the more general case where $Z$ may not be randomized but $X$ is an observed set of baseline covariates that satisfy A1.
To handle the fact that $C$ is only partially observable, additional assumptions are required. A4 is an assumption specific to the identification approach, for example, it can be exclusion restriction or principal ignorability. In the two-sided noncompliance setting we also adopt the common assumption \textit{monotonicity} (A3), which means there are no defiers, leaving only three principal strata.
\begin{alignat*}{2}
    &\text{A0:}~~S=S(Z),~~Y=Y(Z) && \text{(consistency)}
    \\
    &\text{A1:}~~Z\independent (S(1),S(0),Y(1),Y(0))\mid X\quad && \text{(treatment assignment ignorability)}
    \\
    &\text{A2:}~~0<\P(Z=1\mid X)<1 && \text{(treatment assignment positivity)}
    \\
    &\text{A3:}~~S(1)\geq S(0) && \text{(monotonicity, if two-sided noncompliance)}
    \\
    &\text{A4:~~an approach-specific assumption}
\end{alignat*}

Under these assumptions, PCE identification is achieved in three steps. First, the combination of A1, A2 and the $Y$ part of A0 equates PCEs to functions of the distribution of $(X,Z,Y)$ given $C$, which do not involve potential outcomes: 
\begin{align}
    \E[Y(1)-Y(0)\mid C]=\E_{X\mid C}\big\{\E[Y\mid X,Z=1,C]-\E[Y\mid X,Z=0,C]\big\}.\label{eq:unconfoundedness}
\end{align}
Second, A1, A2 and the $S$ part of A0 identify the stratum-specific $X$ distribution over which the outer expectation is taken. (These also identify stratum prevalences, which are usually also of interest.) 
Third, adding A4 (and A3) helps identify the two conditional outcome means, thus identifying the PCE by a function of the joint distribution of $(X,Z,S,Y)$ (and possibly some other observed variables $V$), which involves neither potential outcomes nor $C$.
(Details of this reasoning and step-by-step results are included in Appendix~\ref{asubsec:prelim-aside}%1.1
.)

We will not consider a case involving additional variables $V$ until Section~\ref{sec:recovery}, so we will wait until then to introduce $V$. For now, we proceed with the simple idea that assumptions A0-A4 hold for the PCEs to be equal to some functions of the joint distribution of $(X,Z,S,Y)$. 

This distribution can be factorized as $\P(X,Z,S,Y)=\P(X,Z,S)\P(Y\mid X,Z,S)$, which means if we recover $\P(Y\mid X,Z,S)$ in the presence of outcome missingness, we recover effect identification.
This gives a hint that solving this problem might not need to involve the constructed variable $C$. Our current focus is not to solve this problem from scratch, though, but to examine the LMAR assumption, which involves $C$. We will connect back to this intuition later.

Our discussion of missingness types in Section~\ref{sec:toolbox}-\ref{sec:mar} will use assumptions A0-A1 only, but when recovering effect identification in Section~\ref{sec:recovery} we will use all these assumptions.

\subsection{Missingness types}
\label{subsec:miss-types}

Let $R$ be a binary variable indicating whether $Y$ is observed. 
We start with simple definitions of the missingness types.
LMAR is formally defined as $R\independent Y\mid X,Z,C$, which conditions on the partially observable $C$. Some early writings state this assumption in terms of potential outcomes (e.g., $R(z)\independent Y(z)\mid X,Z=z,C$, where $R(z)$ is potential response), which implies the simpler statement we have here. We use the simpler version, because the job at hand is to handle actual missingness in $Y$, not potential missingness in potential outcomes.
Unlike LMAR, MAR conditions only on observed variables, formally, $R\independent Y\mid X,Z,S$.
The opposite of MAR is MNAR, formally, $R\not\independent Y\mid X,Z,S$.

In the one-sided noncompliance setting, LMAR implies MAR in the treatment arm (where $C$ is observed through $S$) but not in the control arm (where $C$ is unobservable). We refer to these two components of LMAR in this setting as MAR1 and LMAR0.

We will consider settings where $X$ is supplemented by auxiliary variables $W$ in missingness models, so LMAR, MAR and MNAR are defined with $W$ in the conditioning sets. To simplify presentation, we put off the discussion of $W$ until Section~\ref{sec:mar}.

\section{A graphical toolbox}
\label{sec:toolbox}

\subsection{A causal graph to represent principal stratification}
\label{subsec:PS-graph}

To use causal graphs to examine missingness that may depend on principal stratum $C$, we first need a graph that includes $C$.
We will build that graph based on a causal directed acyclic graph (DAG) \citep{pearl1995CausalDiagramsEmpirical}. 
This paper requires very basic knowledge of DAGs: nodes represent variables; arrows represent causal relationships; causal influence flows from upstream to downstream with no circling back; all common causes of any two variables are shown but unobserved unique causes of a variable are typically left off the graph; and the graph encodes no parametric assumption or numerical information. 
With this, we will reason from the ground up.

\begin{figure}[t!]
\centering
\caption{Modifying the DAG to represent the principal stratum variable}
\label{fig:modifying-DAG}
% \resizebox{.75\textwidth}{!}{\input{figures/fig1v2.tkz}}
\includegraphics[width=.75\textwidth]{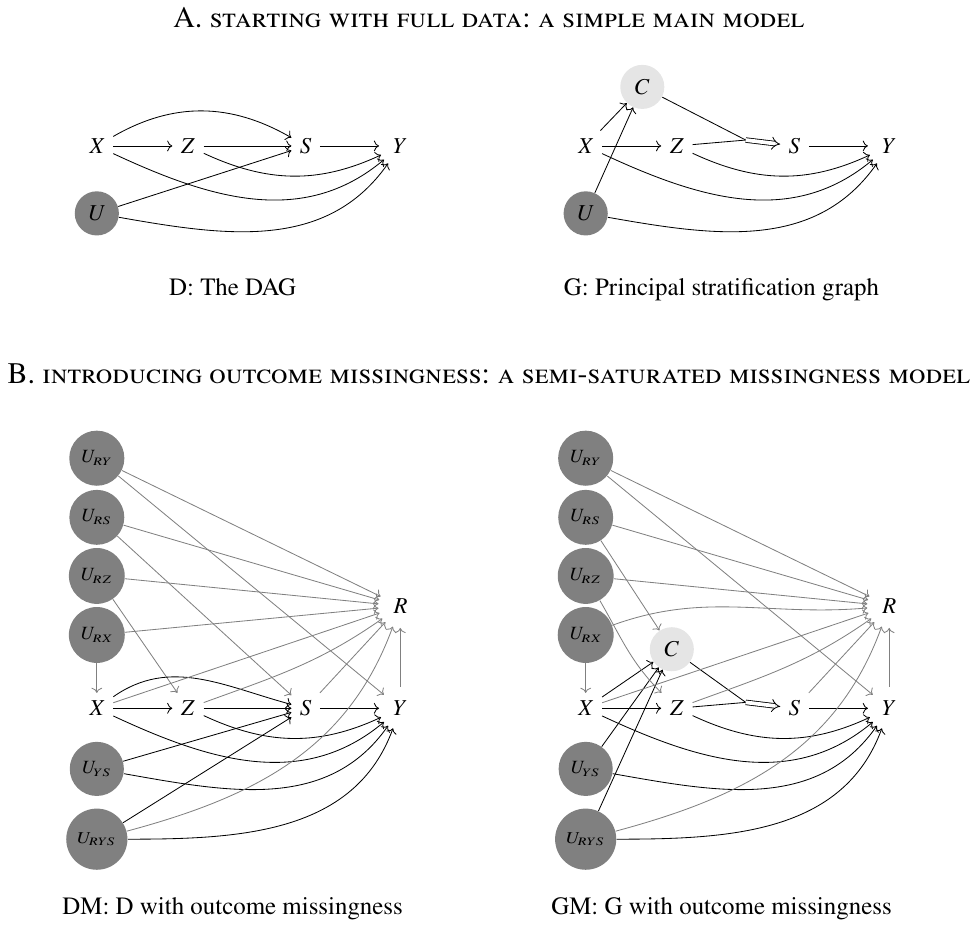}
\end{figure}

We start with a simple DAG labeled D (shown in Figure~\ref{fig:modifying-DAG}A) that respects treatment assignment ignorability, which we call the \textit{simple main model}.
In this model, $X$ captures all common causes that  $Z$ (treatment assigned) shares with $S$ (treatment received) or $Y$ (outcome); this is coherent with treatment assignment ignorability. $S$ and $Y$ may share common causes outside of $X$. These may be observed or unobserved, but for now we put aside the observed and focus on the unobserved common causes.
$U$ represents these unobserved common causes. 
For simplicity, this model does not allow $X$ to share common causes with any of the other variables, as such structures are not central to the consideration of LMAR. Section~\ref{sec:mar} will relax this.

To bring the principal stratum variable $C$ into the graph, we make a small change in the representation of the model for $S$: 
adding $C$ to the graph and (i) have $S$ perfectly determined (indicated by the equal-sign arrow) by $(C,Z)$, and (ii) represent the influence on $S$ of all causes other than $Z$ (including unique causes of $S$ that are implicit in the DAG) as going through $C$. This results in the principal stratification graph labeled G (on the right of Figure~\ref{fig:modifying-DAG}A). 

The equal-sign arrow is borrowed from \cite{shahar2009CausalDiagramsEncoding} with a modification: with multiple causes Shahar has multiple equal-sign arrows going into the perfectly determined variable (one from each cause), whereas we capture the influence of all causes in a single equal-sign arrow.

G is a different type of causal graph from DAGs. DAGs include variables in the observational world; $C$ in graph G is not such a variable, so it behaves differently from other variables in G. The key difference is that, by construction, there are no arrows emitting from $C$ other than the one arrow that goes into $S$. Intuitively, $C$ is the combination of $S(1)$ and $S(0)$, and there is no reason to think that $S(1)$ has an influence on $Y$ in the control arm (where $S(1)$ is not realized) or that $S(0)$ influences $Y$ in the treatment arm (where $S(0)$ is not realized). Any influence of $C$ on $Y$ (or on any other variable in the observational world) has to be through $S$. If $Y$ is dependent on $C$ conditional on observed variables, that must be due to unobserved causes of $C$.

% Any dependence between $C$ and $Y$ conditional on the realized $S$ is therefore due to common causes shared by $Y$ and the counterfactual $S$. Interestingly, this means that such common causes must be pre-treatment-assignment, to be able to influence variables in both worlds.

This is not the only way to represent $C$. An alternative
is to supplement the DAG 
with a component for the counterfactual version of $S$ and have both versions of $S$ combined define $C$. That results in a graph that is more precise in some sense but is more complicated (see Appendix~\ref{asubsec:PS-graph-alternative}%2.2
). We use graph G here because it is simpler and serves our current purpose of considering dependence of outcome and missingness given principal stratum well.

\subsection{Introducing outcome missingness}

We now combine the simple main model with a model for outcome missingness where the missingness may be caused by and/or share common causes with variables in the main model.
The DAG for the combined model (DM, in Figure~\ref{fig:modifying-DAG}B) shows six categories of unobserved common causes, all labeled $U$ with different indices. The four categories $U_{\scriptscriptstyle RX}$, $U_{\scriptscriptstyle RZ}$, $U_{\scriptscriptstyle RS}$, $U_{\scriptscriptstyle RY}$ (placed above $X$) are common causes that $R$ shares with each of the variables $X,Z,S,Y$ alone. These result from allowing the causes of $X$ and the unique causes of $Z,S,Y$ from the main model (implicit in graph D) to influence $R$. The other two categories (placed below $X$) result from considering variables in the original $U$ from graph D and their causes and out of them forming two groups of common causes of $S,Y$: those that do not influence $R$ ($U_{\scriptscriptstyle YS}$) and those that influence $R$ ($U_{\scriptscriptstyle RYS}$). 

We call this missingness model \textit{semi}-saturated for it excludes causes of missingness that may lie on the paths of influence of $X$, $Z$ or $S$ (e.g., on the $X\to Z$ path or on the $Z\to Y$ path). We will consider these \textit{downstream} causes, which are not central to the LMAR discussion, in Section~\ref{sec:mar}.

Similar to the translation from graph D to graph G, we translate graph DM to graph GM (on the right of Figure~\ref{fig:modifying-DAG}B) to include the principal stratum variable $C$. 

In the semi-saturated missingness model, neither LMAR nor MAR holds. In Section~\ref{sec:lmar}, we will consider which restrictions on the model result in LMAR or MAR.

\subsection{Conditional graphs}

To examine missingness assumptions we will use \textit{conditional graphs} derived from causal graphs by conditioning on $X,Z$ and $S$ or $C$. A graph that conditions on a variable taking on a certain value shows (i)~the causal model for the variable's effects specialized to that condition, and (ii)~the distortion of the (in)dependence structure of the variable's causes due to the conditioning.

The conditional graphs in this paper are obtained using three rules: (1)~when conditioning on more than one variable, start with the one most upstream and follow the causal order; (2)~when conditioning on a collider of two variables (a variable caused by them), mark the non-causal dependence induced between them with an undirected dashed edge; (3) drop the variable being conditioned on and all the arrows and edges involving it. Here (2) is self-explanatory, (1) minimizes complexity, and (3) reflects the fact that conditioning on a common cause of two variables removes the dependence (due to the common cause) between them. Side note: the dashed edge in (2) works for our purpose (and is only needed in Section~\ref{sec:mar}), but more complex conditioning (e.g., on a downstream variable but not its causes) requires a more general representation (see Appendix~\ref{asubsec:conditional-general}%2.3
).
% \begin{enumerate}
%     \item when conditioning on more than one variable, start with the one most upstream and then move down the causal hierarchy;
%     \item if the variable being conditioned on has more than one cause, mark the non-causal dependence induced between the causes by adding an undirected dashed edge between each pair of causes;
%     \item drop the variable being conditioned on and all the arrows and edges involving it.
% \end{enumerate}
% The last two rules are based on two facts: conditioning on a collider of two variables (i.e., a variable caused by them) induces dependence between the two causes (known as \textit{collider bias}); and conditioning on a common cause of two variables removes the dependence between these variables due to that common cause.
% Side note: The simple dashed edge works for our purpose (and we will not need it until Section~\ref{sec:mar}), but more complex cases (e.g., conditioning on a downstream variable but not on its causes) require a more general representation (see Appendix~\ref{asubsec:conditional-general}).

\begin{figure}[t!]
\caption{Conditional graphs derived from GM in Figure~\ref{fig:modifying-DAG}}
\label{fig:conditional-graphs}
\centering
% \resizebox{\textwidth}{!}{\input{figures/fig2v2.tkz}}
\includegraphics[width=\textwidth]{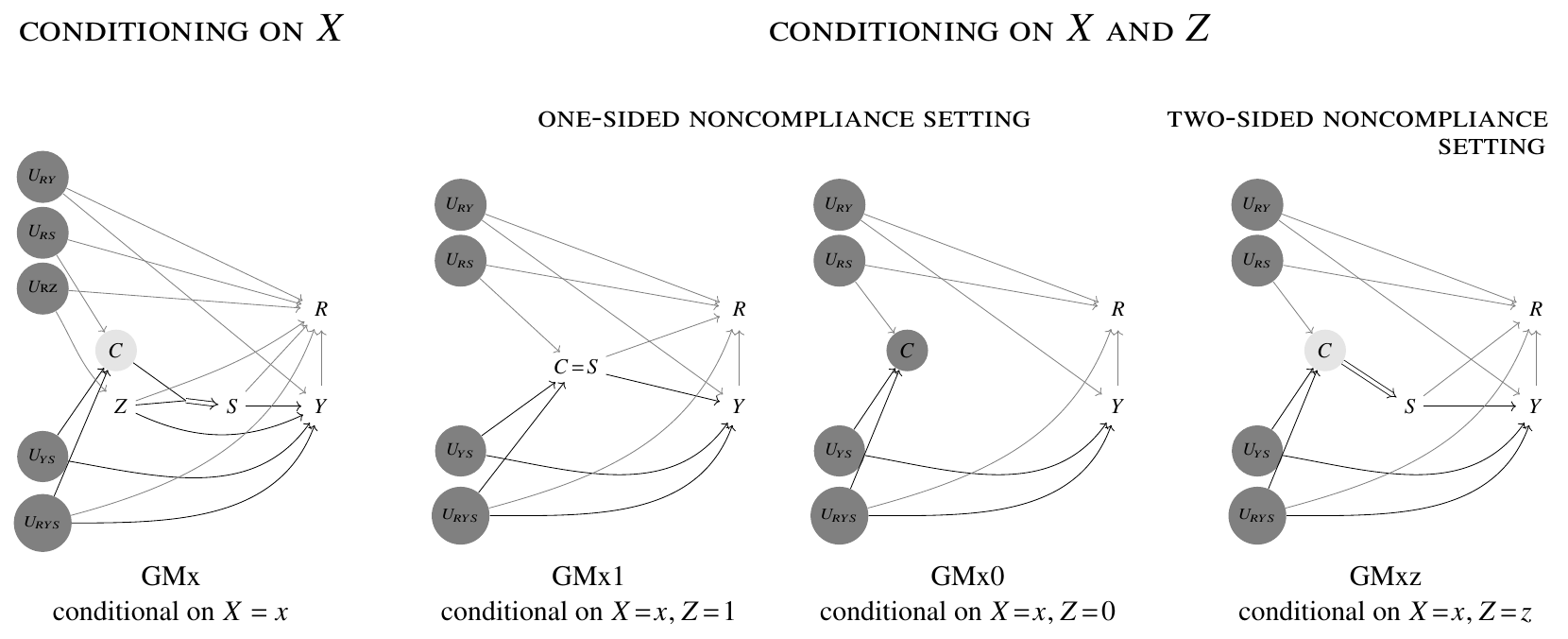}
\end{figure}

In addition, we declutter the derived conditional graphs by (a) dropping any remaining variable that is a constant (with its arrows and edges); (b) combining in one node any pair of adjacent variables that have a one-to-one correspondence; (c) dropping any unobserved variable that has become a unique cause of a variable and is otherwise not connected to the rest of the graph; and (d) dropping any unobserved variable that is not a cause of any other variables on the graph.
% \begin{itemize}
%     \item Order: When conditioning on more than one variable, start with the most upstream one and then move down the causal hierarchy.
%     \item Inducing non-causal dependence: If the variable being conditioned on has more than one cause, add a dashed edge (with no arrows) between each pair of causes to represent the non-causal dependence that is induced between them by the conditioning.
%     \item Breaking dependence: If the conditioning results in a variable being a constant, drop that variable and all the arrows and edges involving that variable.
%     \item Decluttering: If the conditioning turns an unobserved common cause of variables into a unique cause of one variable, drop that cause from the graph.
%     \item Decluttering: If the conditioning results in two adjacent variables having a one-to-one correspondence, combine them in one node on the graph. (This is relevant only because GM has a deterministic relationship.)
% \end{itemize}

Figure~\ref{fig:conditional-graphs} shows conditional graphs derived from graph GM. GMx, which conditions on $X$ taking any value $x$ in its support, is simpler than GM. It does not contain $U_{\scriptscriptstyle RX}$, which has become a unique cause of $R$ when conditioning on $X$. The other graphs, which condition on $Z$ in addition to $X$, are even simpler and do not contain $U_{\scriptscriptstyle RZ}$. Two of these are for the one-sided noncompliance setting. Here $C$ and $S$ have a one-to-one relationship in the treatment arm so they share a node in GMx1; $S$ is a constant (zero) in the control arm so it drops out of GMx0. The last graph, GMxz, applies to either treatment condition in the two-sided noncompliance setting. It looks like GMx1, except $C$ and $S$ remain distinct nodes because their connection is many-to-one.

That two types of $U$ variables drop off the graph when conditioning on $X,Z$ means that with the current simple main model we can ignore them. (With the more general model in Section~\ref{sec:mar} it will be clear that we can ignore $U_{\scriptscriptstyle RZ}$ but not $U_{\scriptscriptstyle RX}$.) For now, we proceed with the simpler model.

\section{Exploration of LMAR and other types of missingness}
\label{sec:lmar}

We consider the model introduced above and three submodels that are more restrictive. For those submodels, we modify the conditional graphs and index them with the letters a, b and c.

\begin{figure}[htbp!]
\caption{Shown in conditional principal stratification graphs: the model from Section~\ref{sec:toolbox} (first graph in each panel) and three submodels (a, b and c), with missingness types indicated}\label{fig:cases}
\centering
\includegraphics[width=\textwidth]{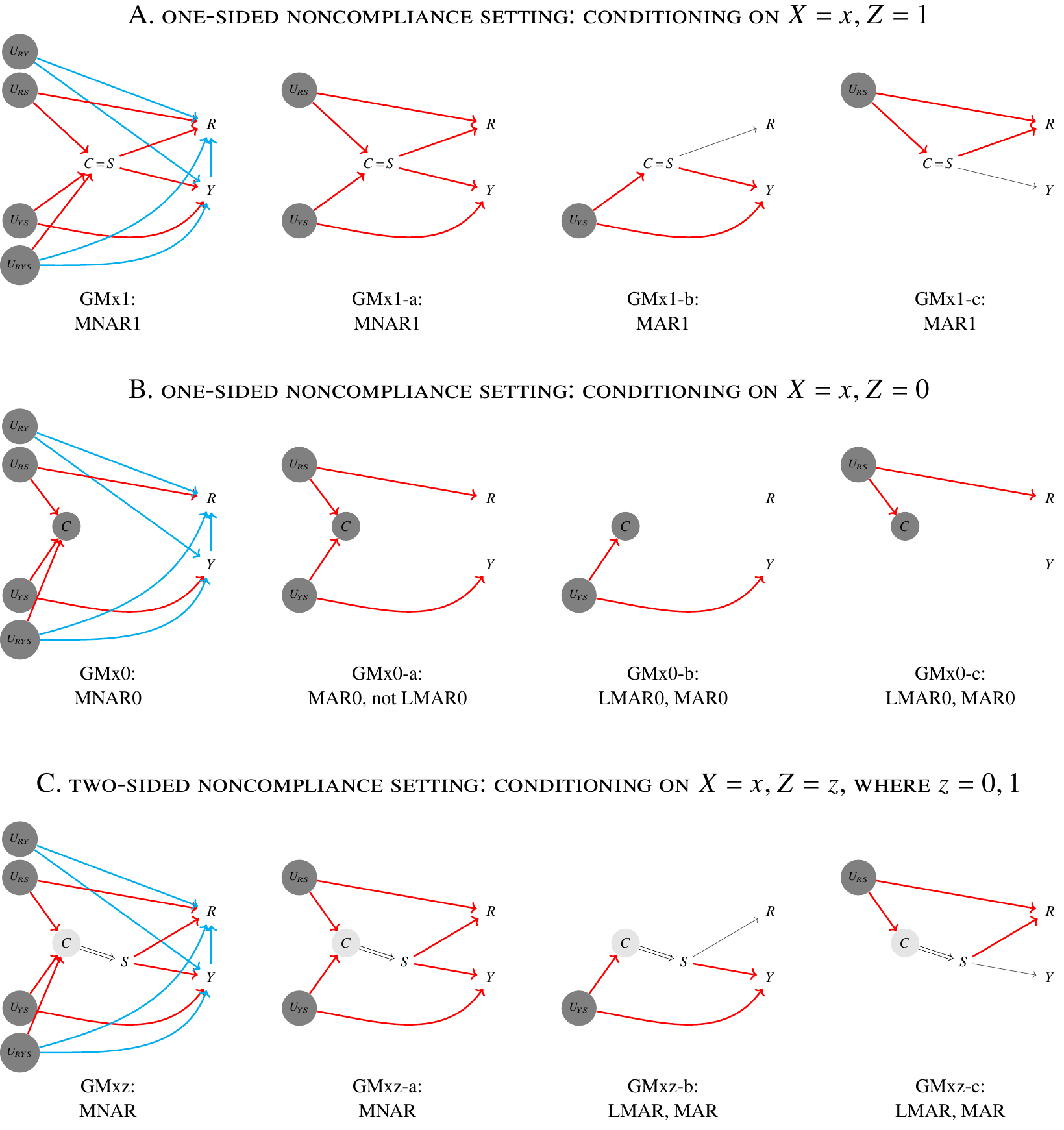}
\end{figure}

\subsection{One-sided noncompliance setting}
\label{subsec:onesided}

Recall that in this setting, LMAR is the combination of MAR1 and LMAR0. First consider the treatment arm (see the top panel of Fig.~\ref{fig:cases}). Graph GMx1, shown again here on the left end, is obviously a MNAR1 case. It has three types of paths through which $R$ is dependent on $Y$:
\begin{enumerate}
    \item the direct path ($Y\rightarrow R$);
    \item paths involving unobserved common causes of $Y$ and $R$ that do not involve $C$ or $S$ ($Y\leftarrow U_{\scriptscriptstyle RYS}/U_{\scriptscriptstyle RY}\rightarrow R$); and
    \item paths that involve $S$ or $C\!=\!S$, including the short path $Y\leftarrow S\rightarrow R$ and the longer paths $Y\leftarrow S\!=\!C\leftarrow U_{\scriptscriptstyle RY}\rightarrow R$ and $Y\leftarrow U_{\scriptscriptstyle YS}/U_{\scriptscriptstyle RYS}\rightarrow C\!=\!S\rightarrow R$.
\end{enumerate}
Next to GMx1 is GMx1-a, which corresponds to a submodel that assumes there are no paths of types 1 and 2. This graph shows what is called a \textit{butterfly} structure \citep{ding2015AdjustNotAdjust} centering the node $C\!=\!S$. Here the type 3 paths can be blocked by conditioning on $S$ (or $C$ in this setting), but such conditioning induces dependence among the two unobserved causes of $C$, which opens up a new path for dependence between $R$ and $Y$, $R\leftarrow U_{\scriptscriptstyle RS}-- U_{\scriptscriptstyle YS}\rightarrow Y$. Hence GMx1-a is also an MNAR1 case. The two more restrictive submodels where either $U_{\scriptscriptstyle RS}$ or $U_{\scriptscriptstyle YS}$ is assumed to be absent (graphs GMx1-b and GMx1-c) break the butterfly structure by removing either the wing involving $R$ or the wing involving $Y$. In these models, the only type 3 path, $R\leftarrow S\rightarrow Y$, can be blocked by conditioning on $S$ (or $C$). GMx1-b and GMx1-c are thus MAR1 cases. 
% And obviously, an even more restrictive model with neither $U_{\scriptscriptstyle RS}$ nor $U_{\scriptscriptstyle YS}$ also satisfies MAR1.

Turning to the control arm (the middle panel of Fig.~\ref{fig:cases}), there are no type 3 paths, but GMx0 is an MNAR0 case due to type 1 and type 2 paths. The submodel removing these paths (GMx0-a) has an M structure, where $R$ and $Y$ are independent unconditional on $C$ but dependent conditional on $C$. Hence GMx0-a satisfies MAR0 but not LMAR0. The two more restrictive submodels that remove one leg of the M (GMx0-b and GMx0-c) satisfy both MAR0 and LMAR0, because in those models $R$ and $Y$ are independent regardless of whether $C$ is conditioned on.

\subsection{Two-sided noncompliance setting}

The bottom panel in Fig.~\ref{fig:cases} concerns the two-sided noncompliance setting, with symmetry between treatment and control arms. The conditional graphs here are of similar forms to the conditional graphs in the top panel. The one difference is that in the current setting conditioning on $C$ is different from conditioning on $S$.
Here GMxz and GMxz-a are both MNAR, whereas GMxz-b and GMxz-c satisfy both MAR and LMAR. With the latter two models, it suffices to condition on the observed variables ($X,Z,S$) to break dependence between $R$ and $Y$; conditioning on $C$ (which has more values than $S$ and is unobserved) is not necessary.

\subsection{MAR versus LMAR and the role of $C$ in missingness assumptions}

The exploration above finds that where LMAR holds MAR also holds, and for one model in one setting, MAR0 holds but LMAR0 does not. This means that contrary to intuition, LMAR is not a relaxation of MAR, but is an assumption of a different kind.
The key takeaway is, for the purpose of rendering $R$ and $Y$ conditionally independent (or reduce the dependence of $R$ and $Y$), it is not necessary to condition on $C$ on top of $X,Z,S$; and conditioning on $C$ may induce unwanted dependence. (This confirms the intuition from Section~\ref{subsec:setting} that recovering $\P(Y\mid X,Z,S)$ needs not involve $C$.) We conclude that $C$ should not be conditioned on in missingness assumptions.

\begin{figure}[t!]
\caption{Same models as in Fig.~\ref{fig:cases}, shown on conditional graphs derived from DAGs}
\label{fig:no-C}
% \resizebox{\textwidth}{!}{\input{figures/fig4v2.tkz}}
\includegraphics[width=\textwidth]{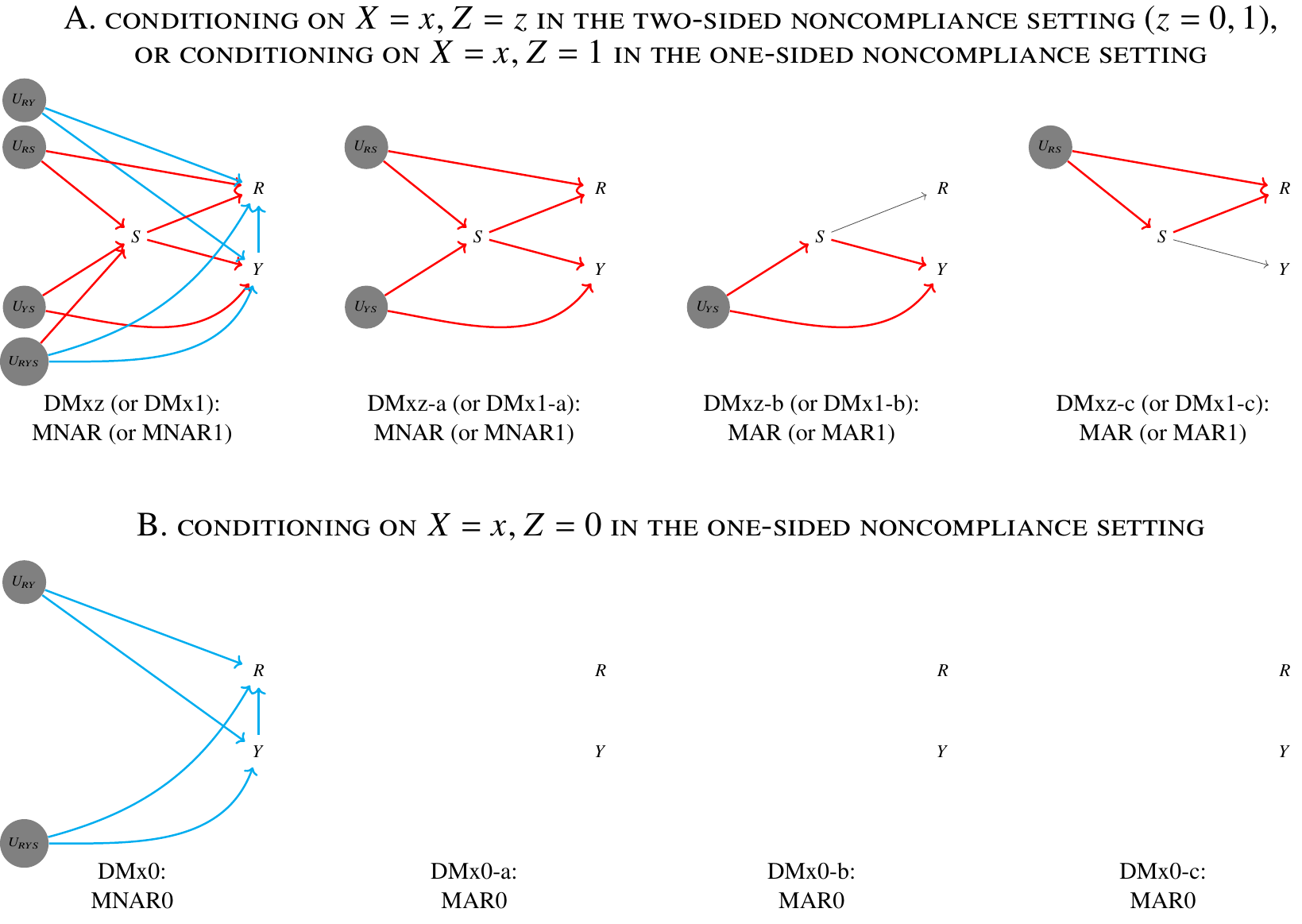}
\end{figure}

We thus revert to using graphs without $C$. Fig.~\ref{fig:no-C} shows the same models using conditional graphs derived from the DAG DM. In the one-sided noncompliance setting, the conditional graphs for the control arm are very simple, and those for the three submodels turn out to be the same graph just with $R$ and $Y$ being (conditionally) independent. This is because $U_{\scriptscriptstyle YS}$ and $U_{\scriptscriptstyle RS}$ are causes of $S$ only in the treatment arm so they do not appear in these graphs.

Also, for the two-sided noncompliance setting, $U_{\scriptscriptstyle RS}$ (or $U_{\scriptscriptstyle YS}$) in DMxz- graphs contains causes that $R$ (or $Y$) shares with $S$, but in GMxz- graphs it contains causes that $R$ (or $Y$) shares with either $S(1)$ and $S(0)$. Hence, at least theoretically, there may be more $U$ variables that get entangled in collider bias when conditioning on $C$ than when conditioning on $S$.

The new insight about MAR versus LMAR suggests that we let go of LMAR and all the specific missingness assumptions that have accompanied LMAR previously, and instead embrace MAR. This is convenient because MAR is simpler to handle and sufficient to recover effect identification (see Section~\ref{sec:recovery}). Also, not requiring the specific missingness assumptions is a plus because they are hard to motivate and some of them (exclusion restriction on response and stable complier response) can conflict with the observed data distribution \citep{nguyen2024IdentificationComplierNoncomplier}. If MAR is deemed unlikely, we should use standard strategies to handle MNAR instead of adopting LMAR.

\section{Further exploration of MAR}
\label{sec:mar}

Now that we have established that MAR should be preferred over LMAR,
this section focuses on MAR. We remove the two key restrictions on the main model and on the missingness model to understand more thoroughly conditions for MAR to hold. We then discuss MAR based on auxiliary variables, to make this assumption more useful in practice.
% Interest in LMAR in practice is probably often motivated by concern that the simpler assumption MAR may not hold; there may be times when making good use of auxiliary variables helps make MAR more realistic. 

% READ THIS AGAIN! Third, we revisit some prior findings involving MAR, LMAR and principal ignorability\todo{I may take this piece out and focus on the final piece.} to clarify any connection between MAR and principal ignorability given our new graphs-based knowledge. This helps ensure, when multiple assumptions need to be invoked for an analysis, that they are coherent with one another. Finally, we derive MAR-based formulas that recover identification of PCEs, where the original identification (in the absence missing data) is based on dominant principal identification approaches. These results serve as a clear starting point for consideration of (existing or new) estimation methods.

\subsection{When $X$ and unobserved causes of $S,Y$ are dependent}
\label{subsec:general-causal}

\begin{figure}[htbp!]
\centering
\caption{When $X$ shares common causes with $S,Y$. On the right, $U_{\scriptscriptstyle RZ}$ and $U_{\scriptscriptstyle XS}$ (which pose no problem across all models) are shown as small gray dots.}
\label{fig:general-main}
% \resizebox{\textwidth}{!}{\input{figures/fig5v2.tkz}}
\includegraphics[width=.9\textwidth]{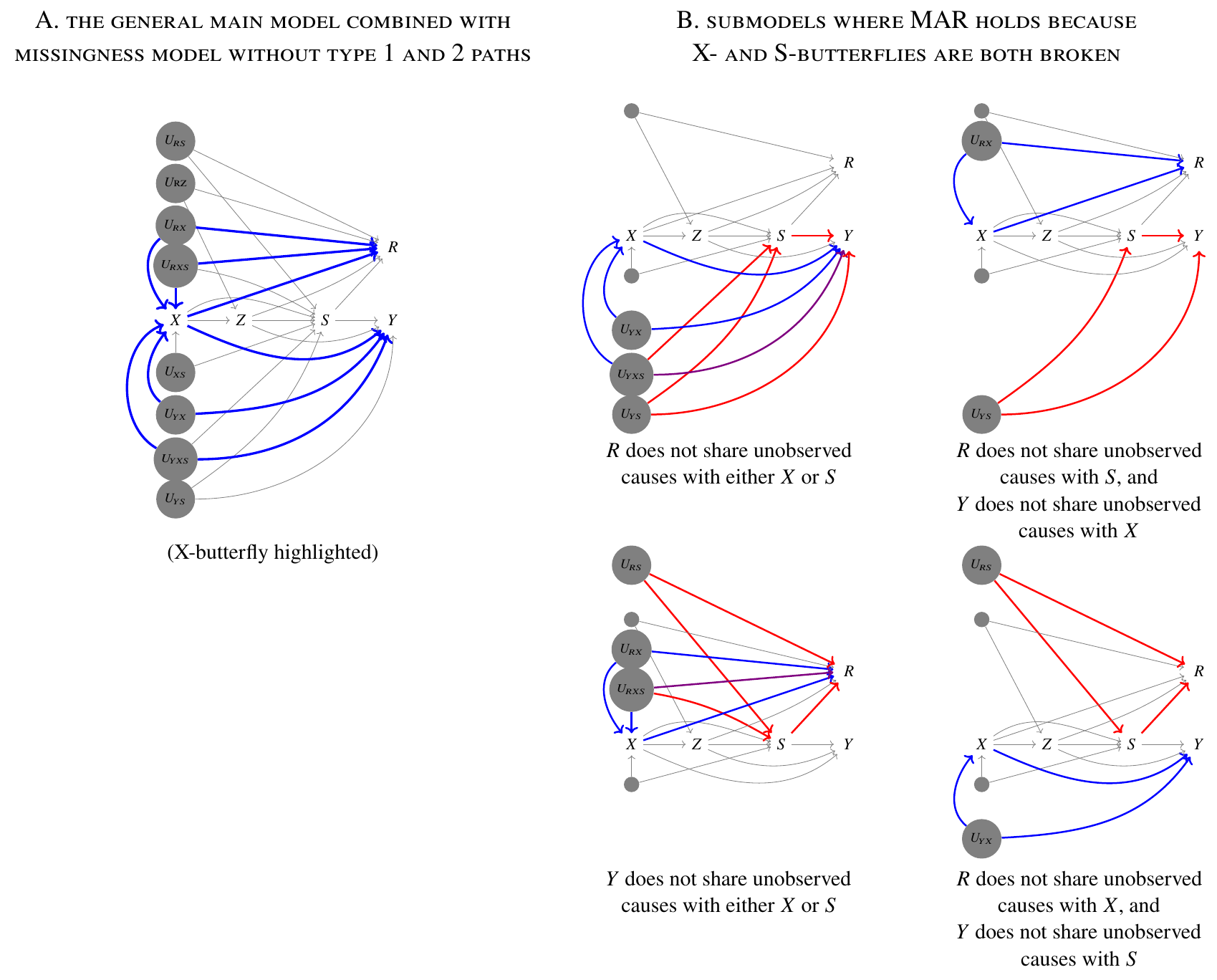}
\caption{When causes of $R$ are downstream. To reduce clutter, all previously considered causes of $R$ are removed, and all unobserved causes of $S,Y$ are collapsed into one node. Unobserved downstream causes of $R$ are represented by small pink dots.}
\label{fig:downstream-U}
% \resizebox{.9\textwidth}{!}{\input{figures/fig6v2.tkz}}
\includegraphics[width=.8\textwidth]{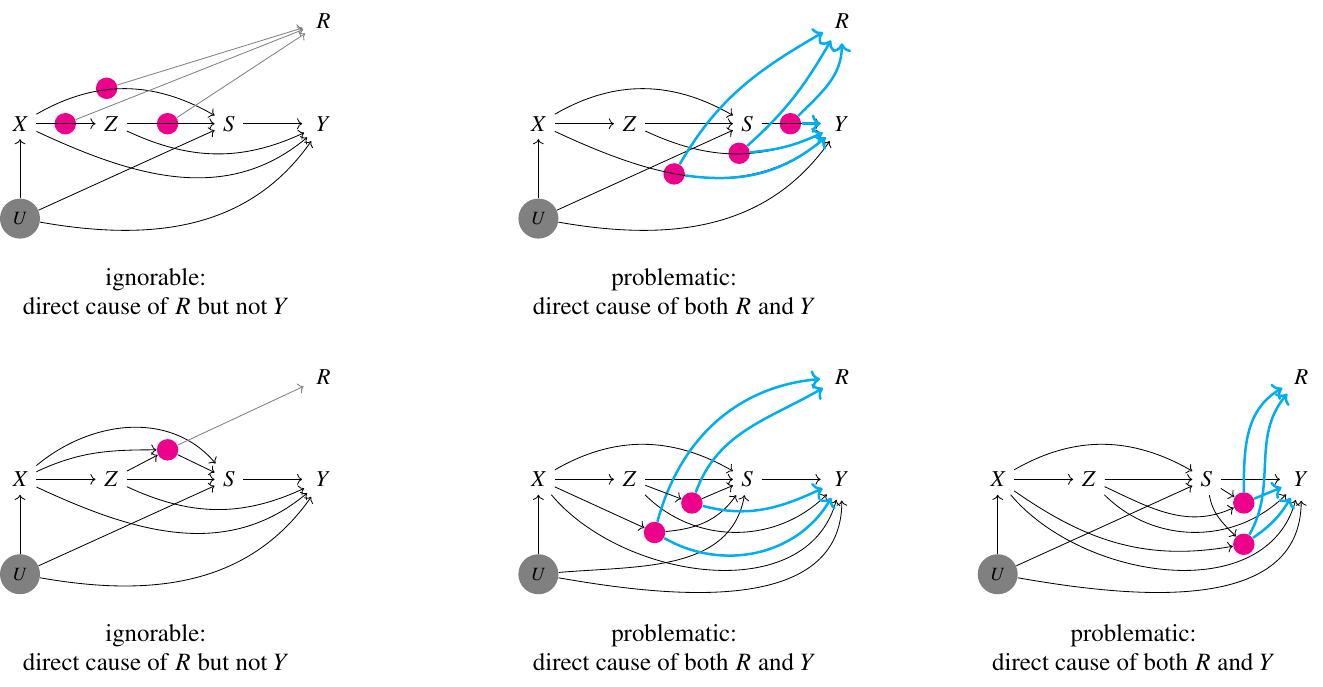}
\end{figure}

The simple main model used so far assumes $X$ is independent of unobserved causes of $Z,S,Y$. This is not necessary to satisfy treatment assignment ignorability though. A model where $X$ shares unobserved causes with $Z$ but not $S,Y$, or a model where $X$ shares unobserved causes with $S$ and/or $Y$ but not $Z$, also satisfies treatment assignment ignorability. We focus on the latter here; the former does not pose any additional conditions required for MAR to hold (see Appendix~\ref{asubsec:third-main-noproblem}%3.1
).

Fig.~\ref{fig:general-main}A shows the general main model combined with the missingness model without paths of types 1 and 2 (to reduce visual clutter). Beside the butterfly structure centering $S$ seen earlier there is now a butterfly structure centering $X$ (highlighted in blue), with wings involving $R$ on one side and wings involving $Y$ on the other side. This structure results in conditional dependence between $R$ and $Y$ -- for both treatment arms and in both the one- and two-sided noncompliance settings -- which violates MAR. A full explanation using graphs that condition on $X,Z,S$ (showing relevant non-causal dependences) is provided in Appendix~\ref{asubsec:general-main}%3.2
.
% 
% This means MAR does not hold if either butterfly structure is present.

Combining results so far, MAR holds if conditions in Box 1 are satisfied.
For the control arm in the one-sided noncompliance setting, $S$ is a constant so (iv) automatically holds. Otherwise, all these conditions are restrictions on the model. (i) and (ii) only concern the missingness model; (iii) and (iv) concern the main and missingness model combined.
Fig.~\ref{fig:general-main}B shows four submodels where these conditions are satisfied, so MAR holds.

\vspace{.5em}

\noindent\fbox{\begin{minipage}{\textwidth}
\small
\textsc{Box 1. Conditions for MAR without auxiliary variables, $R\independent Y\mid X,Z,S$}
\begin{enumerate}[label=(\roman*)]
    \item (\textit{no direct path}) $Y$ does not directly influence $R$;
    \item (\textit{no triangle}) there are no unobserved common causes of $Y$ and $R$;
    \item (\textit{no X-butterfly}) $X$ does not have both unobserved causes shared with $Y$ and unobserved causes shared with $R$;
    \item (\textit{no S-butterfly})\textsuperscript{*} $S$ does not have both unobserved causes shared with $Y$ and unobserved causes shared with $R$.
\end{enumerate}
\footnotesize\textsuperscript{*}\,In the one-sided noncompliance setting, (iv) automatically holds for the control arm, where $S$ is a constant (zero).
\end{minipage}}

\subsection{When causes of missingness are downstream}

We now remove the restriction on the missingness model and let $R$ be influenced by variables downstream of $X,Z,S$. Fig.~\ref{fig:downstream-U} shows possible locations for these causes of $R$ in the main model. Such a cause can be part of one arrow of the main model (see the graphs in the top row in the figure) or part of more than one arrow (the bottom row). The graphs in the first column of the figure shows causes that are ignorable because they are not direct causes of $Y$, so the dependence between $R$ and $Y$ through them are broken when conditioning on $X,Z,S$. The remaining graphs show causes that are not ignorable because they are direct causes of both $Y$ and $R$.

Combining this with what we already know, the four conditions in Box 1 still stand, where (ii) disallows any unobserved common cause of $Y$ and $R$, regardless of its location on the graph.

% \todo{We have been saying no butterfly to mean not both sides of a butterfly ie removing U on one side. But what about no $X/S\to R$ and no $X/S\to M$? That is an M structure where $R$ and $Y$ are independent if we don't condition on $X/S$. That does not recover $\P(Y\mid X,Z,S)$.}

\subsection{MAR with auxiliary variables}
\label{subsec:auxiliary}

The discussion of MAR up to this point has been theoretical. In many applications, it is unlikely that MAR holds conditional on $X,Z,S$ (which was perhaps motivation for the LMAR assumption in the first place). We now expand the discussion of MAR to include auxiliary variables ($W$), i.e., observed variables outside of the set of analysis variables. The hope is MAR holds if $W$ is added to the conditioning set, that is, $R\independent Y\mid X,W,Z,S$. The question is what needs to be assumed about $W$ and about the causal structure for this version of MAR to hold.

\begin{figure}[t!]
\caption{Canonical and instrumental auxiliary variables ($W$)}
\label{fig:auxiliaryW}
% \resizebox{\textwidth}{!}{\input{figures/fig7v2.tkz}}
\includegraphics[width=\textwidth]{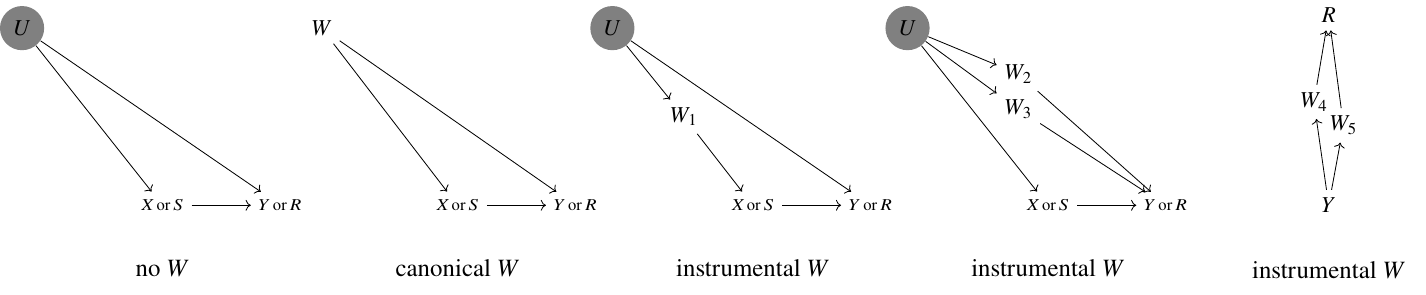}
\end{figure}

First, note that conditions (ii)-(iv) in Box 1 would be satisfied if we could turn those common causes from unobserved to observed (i.e., turn $U$ into $W$). The more common causes are observed, the fewer unobserved ones are left for one to worry about. With such observed common causes (termed \textit{canonical} auxiliary variables), we have a new set of conditions for MAR in Box 2. The first four conditions are directly derived from those in Box 1; the added fifth condition requires that conditioning on $W$ not induce collider bias. As these conditions are simple and clear, they should be used to orient the consideration of auxiliary variables.

\vspace{.5em}

\noindent\fbox{\begin{minipage}{\textwidth}
\small
\textsc{Box 2. Conditions for MAR with canonical auxiliary variables, $R\independent Y\mid X,W,Z,S$}
\begin{enumerate}[label=(\roman*)]
    \item (\textit{no direct path}) $Y$ does not influence $R$;
    \item (\textit{no triangle}) $W$ captures all other common causes of $Y$ and $R$;
    \item (\textit{no X-butterly}) $W$ captures all common causes of $X$ and $Y$ or all common causes of $X$ and $R$;
    \item (\textit{no S-butterfly})\textsuperscript{*} $W$ captures all other common causes of $S$ and $Y$ or all other common causes of $S$ and $R$;
    \item (\textit{no W-butterfly}) no $W$ variable has both unobserved causes shared with $Y$ and unobserved causes shared with $R$.
\end{enumerate}
\footnotesize\textsuperscript{*}\,In the one-sided noncompliance setting, (iv) automatically holds for the control arm, where $S$ is a constant (zero).
\end{minipage}}

\vspace{.5em}

When discussing auxiliary variables in an application, one may notice that some auxiliary variable being considered is not really a common cause of two variables, but is closely related to one. We introduce the concept of \textit{instrumental} auxiliary variables (see Fig.~\ref{fig:auxiliaryW}), as variables on either of the two arrows emitting from an unobserved common cause, which help fully or partially block a path between $Y$ and $R$ that go through the common cause. We also refer to variables that block the otherwise direct path from $Y$ to $R$ as instrumental auxiliary variables. With this broader category of auxiliary variables, we have a new set of conditions for MAR in Box 3.

\vspace{.5em}

\noindent\fbox{\begin{minipage}{\textwidth}
\small
\textsc{Box 3. Conditions for MAR with instrumental/canonical auxiliary variables, $R\independent Y\mid X,W,Z,S$}
\begin{enumerate}[label=(\roman*)]
    \item (\textit{no/controlled direct path})\\$Y$ does not influence $R$;\\or if it does, $W$ blocks that direct path;
    \item (\textit{no/controlled triangles})\\there are no unobserved common causes of $Y$ and $R$;\\or if there are, $W$ blocks all paths between $R$ and $Y$ that go through those unobserved causes;
    \item (\textit{no/controlled X-butterfly})\\$X$ does not have both unobserved causes shared with $Y$ and unobserved causes shared with $R$;\\or if it does, $W$ either blocks all paths between $X$ and $Y$ or blocks all paths between $X$ and $R$ that go through those unobserved causes;
    \item (\textit{no/controlled S-butterfly})\textsuperscript{*}\\$S$ does not have both unobserved causes shared with $Y$ and unobserved causes shared with $R$;\\or if it does, $W$ either blocks all paths between $S$ and $Y$ or blocks all paths between $S$ and $R$ that go through those unobserved causes;
    \item (\textit{no W-butterfly})\\no $W$ variable (or unobserved direct cause of it) has both unobserved causes shared with $Y$ and unobserved causes shared with $R$.
\end{enumerate}
\footnotesize\textsuperscript{*}\,In the one-sided noncompliance setting, (iv) automatically holds for the control arm, where $S$ is a constant (zero).
\end{minipage}}

% \bigskip

% One last note: Above we have referred to the structure that results in collider bias when conditioning on $S$ (or $X$) as a butterfly structure. One may ask, what if $S$ (or $X$) does not influence $Y$ or $R$? In that case, is ``M structure'' a better label? While this is technically correct, in most contexts we expect $S$ (treatment received) to have an effect on $Y$ (outcome), and in many contexts we may not want to assume that $S$ has no effect on $R$. Therefore the \textit{butterfly} label is more general.

\subsection{A brief comment on MNAR}
\label{subsec:mnar}

MNAR requires complicated handling, so it may be helpful to have a sense of how problematic it is in a application. This requires thinking about the causal structure(s) that give rise to MNAR, which is application specific. Here we comment generically on the four relevant types causal structures, as they are not equally problematic.
The most problematic one is the outcome causing its missingness. While theoretically this dependence can be blocked by conditioning on mediators of this causal path, such mediators may not be available (or even exist) except in rare situations.
The second structure, unobserved common causes of $Y$ and $R$ is one that can potentially be mitigated by searching for auxiliary variables that block the dependence.
The X- and S-butterfly structures are in a sense slightly less challenging because they can be mitigated by searching for auxiliary variables that block the paths from the unobserved causes to either $X/S$ or $R/Y$, and not all such paths need to be blocked because one needs to remove only one side of each butterfly structure. Also, even if uncontrolled, due to its indirect nature, dependence due to collider bias from a butterfly structure tends to be less severe than dependence due to a common cause \citep{ding2015AdjustNotAdjust}. This is a rough comparison, however, because the strength of the dependence depends on the strengths of the causal relationships in the structure.

\section{MAR-based recovery of PCE identification}
\label{sec:recovery}

To inform the development of MAR-based methods, we derive identification results for the PCEs under MAR within two dominant effect identification approaches.

\subsection{For the instrumental variable approach}
\label{subsec:iv}

The key assumption of this approach is exclusion restriction. This assumption has appeared in slightly different versions involving either potential or observed outcomes, using either deterministic or stochastic relationships, and either conditional or unconditional on covariates. We use the following version, which is stochastic, conditions on covariates, and importantly, concerns the observed outcome $Y$ and the principal stratum $C=(S(1),S(0))$ -- because we need the assumption to help identify $\E[Y\mid X,Z=z,C]$ in the PCE expression (\ref{eq:unconfoundedness}).
\begin{align*}
    \text{A4a:}~~~Z\independent Y\mid X,S(1)=S(0)~~~~\text{(exclusion restriction)}.
\end{align*}

The DAG in Fig.~\ref{fig:id-assumptions}A shows a main model that satisfies exclusion restriction (for an explanation, see Appendix~\ref{asubsec:exclusion}%4.1
). The difference between this model and our original model is the removal of the arrow from $Z$ to $Y$. Under A4a (and A3), the only non-zero PCE is the CACE, and under full data, it is identified by the well-known \textit{IV formula}:
\begin{align*}
    \text{CACE}=\frac{\E\{\E[Y\mid X,Z=1]-\E[Y\mid X,Z=0]\}}{\E[\P(S=1\mid X,Z=1)-\P(S=1\mid X,Z=0)]}.
\end{align*}
For the connection from (\ref{eq:unconfoundedness}) to this result, see Appendix~\ref{asubsec:iv-identification}%4.3
.
For one-sided noncompliance, the denominator simplifies because $\P(S=1\mid X,Z=0)=0$.

With outcome missingness, we do not observe the conditional outcome mean functions, and need to recover them. Here we use the MAR assumption with auxiliary variables, $R\independent Y\mid X,W,Z,S$. (For MAR without auxiliary variables, $W$ is an empty set.) Since exclusion restriction does not place any restriction on the missingness model, to judge MAR, we need to ask whether conditions in either Box 2 or Box 3 are likely satisfied. 

Using iterated expectation and applying MAR, we obtain
\begin{align*}
    \E[Y\mid X,Z=z]
    &=\E\{\E[Y\mid X,W,Z=z,S,R=1]\mid X,Z=z\}
\end{align*}
for $z=0,1$ in the two-sided noncompliance setting and for $z=1$ in the one-sided noncompliance setting.
For $z=0$ in the one-sided noncompliance setting, this is replaced with
\begin{align*}
    \E[Y\mid X,Z=0]=\E\{\E[Y\mid X,W',Z=0,R=1]\mid X,Z=0\},
\end{align*}
where $W'$ is a subset of $W$ removing variables that only serve the purpose of breaking the S-butterfly structure under $Z=1$.

\subsection{For the principal ignorability approach}

Principal ignorability has often been stated in the literature as an assumption about potential outcomes, but we follow \cite{nguyen2024IdentificationComplierNoncomplier} and state it as an assumption about the observed outcome $Y$. Again, this is motivated by the fact that we need the assumption to help identify $\E[Y\mid X,Z=z,C]$ in the PCE expression in (\ref{eq:unconfoundedness}). The statement of the assumption in this prior work is limited to the one-sided noncompliance setting and uses the same covariate set as that in the treatment assignment ignorability assumption ($X$). Here we generalize the assumption statement so that it applies to both settings and allows conditioning on a larger set of baseline covariates ($X,V$):
\begin{align*}
    \text{A4b:}~~~Y\independent C\mid X,V,Z,S~~~~\text{(general principal ignorability)}.
\end{align*}

The idea of this assumption is that, within each mixture of principal strata in a study arm (e.g., the mixture of compliers and always-takers seen with $Z=1,S=1$), conditional on covariates ($X,V$), principal stratum is independent of the outcome.
For one-sided noncompliance, A4b reduces to $Y\independent C\mid X,V,Z=0$. Also, $V$ can be the empty set if that is reasonable in an application.

Fig.~\ref{fig:id-assumptions}B shows two main models that satisfy general principal ignorability (see explanation in Appendix~\ref{asubsec:principal-ignorability}%4.2
), with two characteristics: (i)~$X$ and $V$ capture all common causes of $S,Y$, and (ii)~$X$/$V$ can have unobserved causes shared with $S$ or with $Y$ (hence the two models) but not both. 

Under A4b (and A3), in the two-sided noncompliance setting, the effects are identified as
\begin{align*}
    \text{CACE}&=\E\left(\big\{\E[Y\mid X,V,Z=1,S=1]-\E[Y\mid X,V,Z=0,S=0]\big\}\frac{p_1(X,V)-p_0(X,V)}{\E[p_1(X,V)-p_0(X,V)]}\right),
    \\
    \text{NACE}&=\E\left(\big\{\E[Y\mid X,V,Z=1,S=0]-\E[Y\mid X,V,Z=0,S=0]\big\}\frac{1-p_1(X,V)}{\E[1-p_1(X,V)]}\right),
    \\
    \text{AACE}&=\E\left(\big\{\E[Y\mid X,V,Z=1,S=1]-\E[Y\mid X,V,Z=0,S=1]\big\}\frac{p_0(X,V)}{\E[p_0(X,V)]}\right),
\end{align*}
where $p_z(X,V):=\P(S=1\mid X,V,Z=z)$. In the one-sided noncompliance setting, 
\begin{align*}
    \text{CACE}&=\E\left(\big\{\E[Y\mid X,V,Z=1,S=1]-\E[Y\mid X,V,Z=0]\big\}\frac{p_1(X,V)}{\E[p_1(X,V)]}\right),
    \\
    \text{NACE}&=\E\left(\big\{\E[Y\mid X,V,Z=1,S=0]-\E[Y\mid X,V,Z=0]\big\}\frac{1-p_1(X,V)}{\E[1-p_1(X,V)]}\right).
\end{align*}
These results (see proof in Appendix~\ref{asubsec:pi-identification}%4.4
) are similar to results in \cite{jiang2022MultiplyRobustEstimationa} and \cite{nguyen2024SensitivityAnalysisPrincipal}, except for the presence of $V$.

\begin{figure}[t!]
\caption{Bringing assumption A4 into the DAG}
\label{fig:id-assumptions}
% \resizebox{\textwidth}{!}{\input{figures/v2tkz/fig8v2.tkz}}
\includegraphics[width=\textwidth]{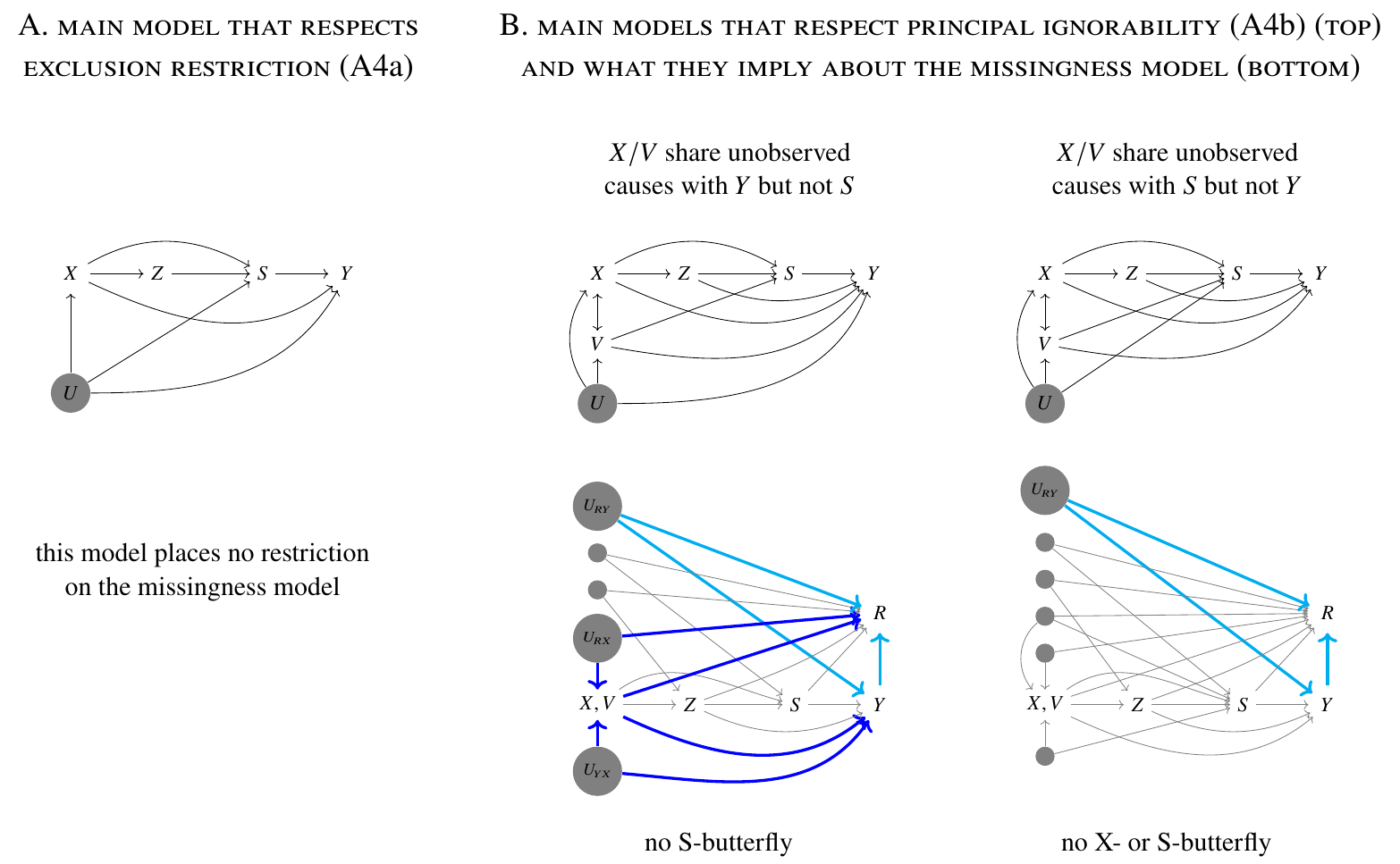}
\end{figure}

With outcome missingness, we need to recover the conditional outcome mean functions. The MAR assumption with auxiliary variables is $R\independent Y\mid X,V,W,Z,S$.
Unlike exclusion restriction, the principal ignorability models here place restrictions on the missingness model. Because there are no unobserved common causes of $S,Y$, there is no S-butterfly structure, which means when considering the plausibility of MAR, we can skip condition (iv). Additionally, if $X,V$ are believed to not share unobserved common causes with $Y$, there is also no butterfly structure centering $X,V$.

Again, using iterated expectation and then applying MAR, we have
\begin{align*}
    \E[Y\mid X,V,Z=z,S=s]=\E\{\E[Y\mid X,V,W,Z=z,S=s,R=1]\mid X,V,Z=z,S=s\}
\end{align*}
for $z=0,1$ in the two-sided noncompliance setting and $z=1$ in the one-sided noncompliance setting, and for $s=0,1$.
In the one-sided noncompliance setting, we also have
\begin{align*}
    \E[Y\mid X,V,Z=0]=\E\{\E[Y\mid X,V,W,Z=0,R=1]\mid X,V,Z=0\}.
\end{align*}

\section{Discussion}
\label{sec:conclusion}

This paper shows that MAR is a weaker assumption than LMAR and should be preferred over LMAR in handling outcome missingness in principal stratification analysis. It then clarifies conditions on the causal structure and auxiliary data for MAR to hold, and uses MAR to recover effect identification for two major effect identification approaches. It highlights the structures that give rise to MNAR in this setting. With this work, we hope to help strengthen the foundations on which missing data methods are built.

We comment on a few related topics. First is a practical question: based on our results about MAR, which variables should be used (at analysis stage), or collected (at study design stage), to serve as auxiliary variables in handling outcome missingness. We recommend to think (at both these stages) about causes of the outcome, causes of its missingness, and also causes of the post-treatment variable (treatment received). Good auxiliary variables are common causes of two types: those shared by $Y$ and $R$, and those that $Y$ or $R$ shares with $S$ or $X$ (and particularly $S$). Proxies of such causes may be used, but require careful examination of the causal structure.

Second, while this work focuses on principal stratification analysis targeting principal causal effects (effects of treatment assignment within principal strata), the conditions for MAR to hold in Section~\ref{sec:mar} are relevant to the classic instrumental variable analysis that targets the local average treatment effect (LATE, effect of treatment received on the compliers), because while the effect definitions differ, the setting is the same. Also, the CACE recovery result within the instrumental variable approach in Section~\ref{subsec:iv} is also relevant to the LATE, because under the typical LATE identification assumptions, LATE and CACE are equal. 

Third, as conditional independence assumptions are often used to handle complexities in analysis, we note that the finding about LMAR here is relevant to any other assumption that \textit{conditions on principal stratum}. In generic terms, the finding is two-fold: (i) conditioning on $C$ (on top of $S$) does not remove dependence among realized variables because $C$ is not a direct cause of such variables; and (ii) conditioning on $C$ induces non-causal dependence among all causes of $S(1)$ and $S(0)$, which might be problematic if they are causally connected to the variables one wishes to assume to be conditionally independent.
These issues do not affect assumptions about principal stratum that do not condition on principal stratum, e.g., principal ignorability.

Fourth, after working with this setting for some time, we have arrived at a general conclusion that it helps to deliberately use one assumption for each distinct task. In this problem, this helps pinpoint what is needed of an assumption. Specifically, after invoking treatment assignment ignorability, the approach-specific identification assumption is needed to disentangle the dependence of the realized outcome (not potential outcome) on $C$; this is something we became aware of in \cite{nguyen2024IdentificationComplierNoncomplier} but not in our earlier paper \cite{nguyen2024SensitivityAnalysisPrincipal}. And then the missingness assumption is needed to recover the conditional distribution/mean of the realized outcome conditional on observed variables (not $C$); we noticed this here but not in \cite{nguyen2024IdentificationComplierNoncomplier}. We believe that this one-assumption-per-task approach may be generally fruitful with complex problems.

Last but not least, causal graphs are powerful tools. As with any tool, advanced knowledge may help accomplish very complex tasks, while basic knowledge (like what we used) can help with less complex but important tasks. As causal inference and missing data methods rely heavily on assumptions, we hope there will be more of this kind of simple use of causal graphs to critically examine those assumptions.

\section*{Acknowledgement}
This work is supported by grants R03MH128634 (PI Nguyen) and U24OD023382 (PI Jacobson) from the US National Institutes of Health. The author appreciates helpful discussions from colleagues at the Stuart Lab at Johns Hopkins Bloomberg School of Public Health.

\bibliography{refs}

\vspace{4em}

% \clearpage
% \pagenumbering{arabic}% resets `page` counter to 1
% \renewcommand*{\thepage}{A\arabic{page}}
\appendix
{\Large \textbf{APPENDIX}}

\vspace{-5em}

\part{}
\parttoc

\renewcommand\thefigure{S\arabic{figure}}    
\setcounter{figure}{0}

\section{Appendix to Section~\ref{sec:prelim}}
\label{asec:prelim}

\subsection{PCE identification steps}
\label{asubsec:prelim-aside}

Section~\ref{subsec:principal-assumptions} states that PCE identification -- in the absence of missing data -- is achieved in three steps. Here are the steps in details.

\textbf{Step 1}: Use assumptions A1, A2 and the $Y$ part of A0 to establish that 
\begin{align*}
    \E[Y(1)-Y(0)\mid C]=\E_{X\mid C}\{\E[Y\mid X,Z=1,C]-\E[Y\mid X,Z=0,C]\}
    \tag{\ref{eq:unconfoundedness}}
\end{align*}

The key is to show that $Z\independent(Y(1),Y(0))\mid X,C$. We will use the lemma: $A\independent(B,C)\Longrightarrow A\independent B\mid C$. To prove this lemma, first note that $A\independent(B,C)\Longrightarrow A\independent C$:
\begin{align*}
    \P(A\mid C)
    &=\E[\P(A\mid B,C]\mid C] && (\text{law of total probability})
    \\
    &=\E[\P(A)\mid C] && (A\independent(B,C))
    \\
    &=\P(A).
\end{align*}
Then $A\independent B\mid C$ follows, because
\begin{align*}
    \P(B\mid C)
    &=\frac{\P(B,C)}{\P(C)} && (\text{Bayes' rule})
    \\
    &=\frac{\P(B,C\mid A)}{\P(C\mid A)} && (\text{because}~A\independent(B,C)~\text{and}~A\independent C)
    \\
    &=\P(B\mid C,A). && (\text{Bayes' rule})
\end{align*}

Applying this lemma, it follows from assumption A1 (treatment assignment ignorability) that
\begin{align*}
    Z\independent(Y(1),Y(0))\mid X,C.
\end{align*}

Now we use the standard causal inference reasoning: for $z=0,1$
\begin{align*}
    \E[Y(z)\mid C]
    &=\E\{\E[Y(z)\mid X,C]\mid C\} && (\text{iterated expectation})
    \\
    &=\E\{\E[Y(z)\mid X,C,Z=z]\mid C\} && (Z\independent Y(z)\mid X,C)
    \\
    &=\E\{\E[Y\mid X,C,Z=z]\mid C\}. && (\text{positivity (A2) and consistency (the $Y$ part of A0)})
\end{align*}

\textbf{Step 2}: Identify $\P(X\mid C)$ using A1, A2 and the $S$ part of A0. 

We start with
\begin{align*}
    \P(X\mid C)
    &=\P(X)\frac{\P(X\mid C)}{\P(X)}
    \\
    &=\P(X)\frac{\P(C\mid X)}{\P(C)} && (\text{Bayes' rule})
    \\
    &=\P(X)\frac{\P(C\mid X)}{\E[\P(C\mid X)]} && (\text{total probability})
    \\
    &=\P(X)\frac{\P(C\mid X,Z=z)}{\E[\P(C\mid X,Z=z)]}, && (\text{treatment assignment ignorability A1)}
\end{align*}
for $z=0,1$.

In the one-sided noncompliance setting,
\begin{align*}
    \P(\text{complier}\mid X,Z=1)
    &=\P(S(1)=1\mid X,Z=1) && (\text{by definition})
    \\
    &=\P(S=1\mid X,Z=1), && (\text{positivity (A2) and consistency (the $S$ part of A0)})
\end{align*}
and similarly,
$\P(\text{noncomplier}\mid X,Z=1)
    =
    \P(S=0\mid X,Z=1)$.
It follows that
\begin{alignat*}{3}
    &\P(X\mid\text{complier})
    &&=\P(X)\frac{\P(S=1\mid X,Z=1)}{\E[\P(S=1\mid X,Z=1)]},
    \\
    &\P(X\mid\text{noncomplier})
    &&=\P(X)\frac{\P(S=0\mid X,Z=1)}{\E[\P(S=0\mid X,Z=1)]}.
\end{alignat*}

In the two-sided noncompliance setting,
\begin{align*}
    \P(\text{always-taker}\mid X,Z=0)
    &=\P(S(0)=1\mid X,Z=0) && (\text{monotonicity (A3)})
    \\
    &=\P(S=1\mid X,Z=0), && (\text{positivity (A2) and consistency (the $S$ part of A0)})
\end{align*}
and similarly, $\P(\text{never-taker}\mid X,Z=1)=\P(S=0\mid X,Z=1)$. And
\begin{align*}
    \P(\text{complier}\mid X,Z=1)
    &=\P(\text{complier or always-taker}\mid X,Z=1)-\P(\text{always-taker}\mid X,Z=1)
    \\
    &=\P(S(1)=1\mid X,Z=1)-\P(\text{always-taker}\mid X,Z=1)~~~(\text{by definition})
    \\
    &=\P(S=1\mid X,Z=1)-\P(\text{always-taker}\mid X,Z=1)~~~(\text{positivity and consistency})
    \\
    &=\P(S=1\mid X,Z=1)-\P(\text{always-taker}\mid X,Z=0)~~~(\text{treatment assignment ignorability})
    \\
    &=\P(S=1\mid X,Z=1)-\P(S=1\mid X,Z=0).
\end{align*}
It follows that
\begin{alignat*}{3}
    &\P(X\mid\text{complier})
    &&=\P(X)\frac{\P(S=1\mid X,Z=1)-\P(S=1\mid X,Z=0)}{\E[\P(S=1\mid X,Z=1)-\P(S=1\mid X,Z=0)]},&&
    \\
    &\P(X\mid\text{always-taker})
    &&=\P(X)\frac{\P(S=1\mid X,Z=0)}{\E[\P(S=1\mid X,Z=0)]},&&
    \\
    &\P(X\mid\text{never-taker})
    &&=\P(X)\frac{\P(S=0\mid X,Z=1)}{\E[\P(S=0\mid X,Z=1)]}.&&
\end{alignat*}

Side note: The derivation above also yields the prevalences of the principal strata, which are
\begin{alignat*}{3}
    &\P(\text{complier})&&=\E[\P(S=1\mid X,Z=1)],&&
    \\
    &\P(\text{noncomplier})&&=\E[\P(S=0\mid X,Z=1)]&&
\end{alignat*}
for the one-sided noncompliance setting, and
\begin{alignat*}{3}
    &\P(\text{complier})&&=\E[\P(S=1\mid X,Z=1)-\P(S=1\mid X,Z=0)],&&
    \\
    &\P(\text{always-taker})&&=\E[\P(S=1\mid X,Z=0)],&&
    \\
    &\P(\text{never-taker})&&=\E[\P(S=0\mid X,Z=1)]&&
\end{alignat*}
for the two-sided noncompliance setting.

\textbf{Step 3}: Use the approach-specific assumption A4 to identify the stratum-specific conditional outcome means $\E[Y\mid X,Z=z,C]$, which are part of the formula (\ref{eq:unconfoundedness}).

We leave this step to Section~\ref{sec:recovery} of the paper and Section~\ref{asec:recovery} of the Appendix.

\section{Appendix to Section~\ref{sec:toolbox}}
\label{asec:toolbbox}

\subsection{Connecting main model DAGs and treatment assignment ignorability}
\label{asubsec:treatment-assignment-ignorability}

Sections~\ref{sec:toolbox} and \ref{sec:lmar} in the paper uses the \textit{simple main model}. Section~\ref{sec:mar} considers the \textit{general main model} and mentions a third model where $X$ shares unobserved common causes with $Z$ but not $S,Y$ (which is examined in detail in Appendix~\ref{asubsec:third-main-noproblem}). Here we show that these models satisfy treatment assignment ignorability.

\begin{figure}[h!]
\caption{Three main models that satisfy treatment assignment ignorability, $Z\independent(S(1),S(0),Y(1),Y(0))\mid X$ shown in DAGs (top row), SWIGs (middle row) and both-worlds graphs (bottom row)}
\label{fig:treatment-assignment-ignorability}
\resizebox{\textwidth}{!}{\input{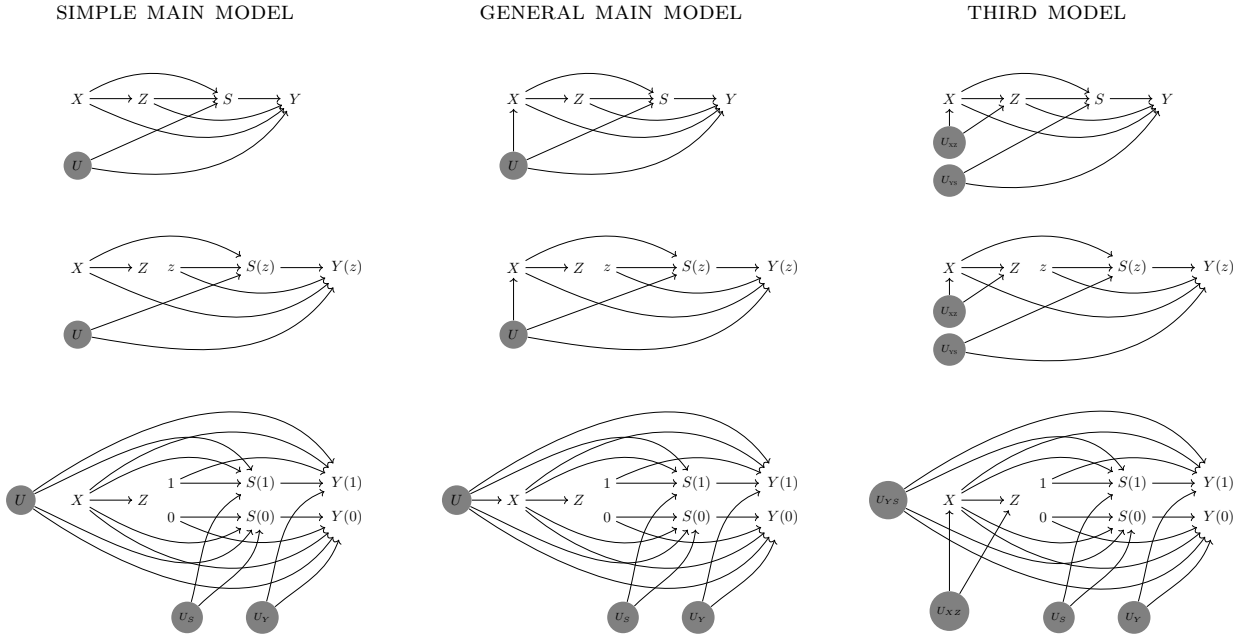}}
\end{figure}

We start with the DAGs for these three models in the top panel of Figure~\ref{fig:treatment-assignment-ignorability}, and proceed in two steps. First, consider the single-world interventional graphs (SWIGs) \citep{richardson2013SingleWorldIntervention} based on these DAGs (middle panel of Figure~\ref{fig:treatment-assignment-ignorability}). These graphs reveal the relationship of the potential treatment received $S(z)$ and potential outcome $Y(z)$ (for $z=0,1$) with other variables. All of them show that $Z\independent(S(z),Y(z))\mid X$, so we are almost there. Second, to fully connect to the treatment assignment ignorability assumption A1, $Z\independent(S(1),S(0),Y(1),Y(0))\mid X$, we use the both-worlds graphs (bottom panel of Figure~\ref{fig:treatment-assignment-ignorability}). These graphs look complicated, but each one is really just the combination of the two $z=1$ and $z=0$ SWIGs. The additional $U$ variables are just the unique causes $U_{\scriptscriptstyle S}$ of $S$ and $U_{\scriptscriptstyle Y}$ of $Y$ (which are implicit in the DAGs and SWIGs) being made explicit. All these graphs show $Z\independent(S(1),S(0),Y(1),Y(0))\mid X$.

\subsection{The alternative principal stratification graph}
\label{asubsec:PS-graph-alternative}

\begin{figure}[h]
\caption{Representations of principal stratification -- based on the general main model}
\label{fig:PS-graph-constructive}
\resizebox{\textwidth}{!}{\input{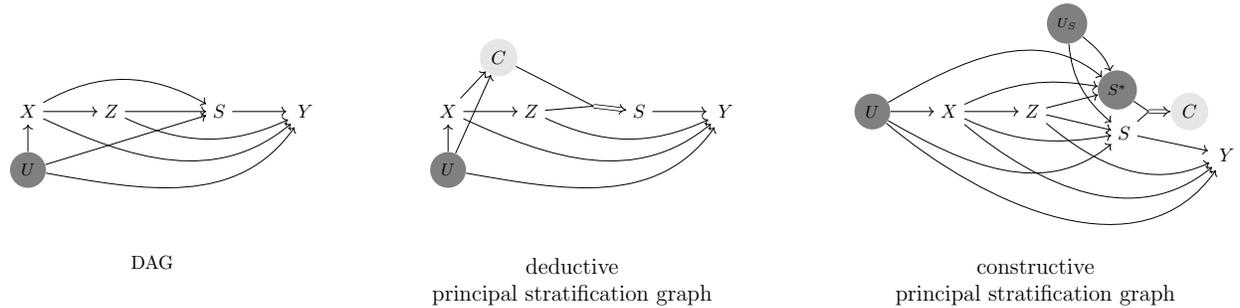}}
\end{figure}

In Section~\ref{subsec:PS-graph} in the paper we constructed a principal stratification graph, which we will call \textit{deductive}, reflecting that we ``deduce'' $S$ from its potentials ($C=(S(1),S(0))$). We use that graph in the paper because it is simple and it fits with the usual idea that the observed outcome reveals a potential outcome. We also mentioned an alternative graph for principal stratification, which we will call \textit{constructive}, reflecting that we ``construct'' the principal stratum variable $C$. Figure~\ref{fig:PS-graph-constructive} shows this alternative graph next to the deductive one.

The key difference between the two representations is that the constructive graph allows considering the causes of $S(1)$ and $S(0)$ separately (if necessary) while the deductive graph lumps them all together as causes of $C$.
% the constructive graph keeps the causes of the factual variable $S$ directly connected to $S$ (while the deductive graph have those causes influence $S$ via influencing $C$). This feature allows considering the causes of $S(1)$ and of $S(0)$ separately, whereas the deductive graph lumps them all together in the group of causes of $C$. 
For the problems we tackle, the two representations do not lead to results that are qualitatively different. We use the deductive graph in the paper because it is simpler.

% This feature can be important when considering the relationship between $C$ and other variables when conditioning on $S$. (For example, when evaluating principal ignorability, formally, $C\independent Y\mid X,Z,S$, if $Y$ shares unobserved causes with $S$ but not with its counterfactual $S^*$, a constructive graph will correctly inform that principal ignorability holds, but a deductive graph will not because it lumps all the non-$X$ causes of $S$ and $S^*$ together.) For our central purpose of evaluating LMAR (importantly whether $R$ and $Y$ are independent conditional on $C$) rather than whether $C$ is independent of some variable when conditioning on $S$, the deductive graph suffices, and is simpler.

\subsection{Representation of noncausal dependence induced by conditioning}
\label{asubsec:conditional-general}

In the paper, when conditioning on a collider between more than one cause, we add (to the conditional graph) a dashed edge between each pair of causes to represent the noncausal dependence that is induced by conditioning. All this says is that the conditional joint distribution of the causes is different from their unconditional joint distribution. This dashed edge representation works well for our conditioning act, which conditions on the combination of $X,Z,S$ where $X$ is the first observed variable in the causal structure, and $Z,S$ follow immediately after $X$. In a different situation where one wishes to condition on a downstream variable without conditioning on its causes, e.g., conditioning on $S=s$ only but not on $X,Z$, then this dashed edge representation is not ideal. The conditioning act changes the joint distribution of all the causes, so we would need $m$-choose-2 dashed edge (for $m$ causes), in addition to all the causal arrows, which would clutter the graph very quickly. Rather than marking pairwise dependence in that case, it is better to mark all the causes of the variable being conditioned on as a group. All variables in that group now have noncausal dependence with one another, and any two variables outside of the group that have respective causes in the group are dependent in the condition.

\begin{figure}
\caption{General representation of induced noncausal dependence in a conditional graph: simply marking the causes of the variable being conditioned on and the relevant causal links. The example here is the conditional graph given $S=s$ based on graph DM in Figure~\ref{fig:modifying-DAG}B, with the same \textit{collider fan} shown in different ways.}
\label{fig:colliderfan}
\resizebox{\textwidth}{!}{\input{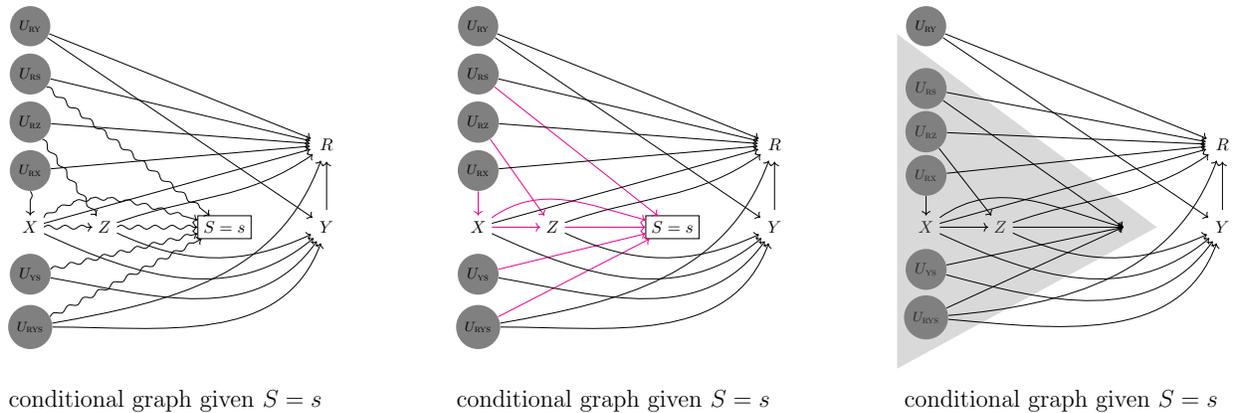}}
\end{figure}

Figure~\ref{fig:colliderfan} shows the conditional graph, with several (among many) ways that the group of causes can be marked. In this case, $S$ has seven causes, five direct and two indirect. The dashed edge representation would require 21 dashed edges, which would be unmanageable. The marking as general representation does not add edges, but rather, uses color or line shape, or background shading to convey the same information. Due to its shape on the graph, we call the group of causes being marked the \textit{collider fan} of $S$.

With this representation, we can see that conditional on $S=s$, $R$ and $Y$ are dependent through two paths that are completely outside the collider fan and through many paths that involve variables in the collider fan. We can also continue to use information from the causal structure as with the unconditional graph. For example, we can see that if one additionally conditions on $X$, that blocks the causal path that links $U_{\scriptscriptstyle RX}$ to the rest of the collider fan, so it releases this node from the collider fan.

\section{Appendix to Section~\ref{sec:mar}}
\label{asec:mar}

\subsection{The case where $X$ shares unobserved causes with $Z$ but not with $S,Y$}
\label{asubsec:third-main-noproblem}

\begin{figure}[h!]
\caption{The third main model from Figure~\ref{fig:treatment-assignment-ignorability} combined with the simple missingness model minus type 1 paths, shown in unconditional and conditional graphs}
\label{fig:third-main}
\resizebox{\textwidth}{!}{\input{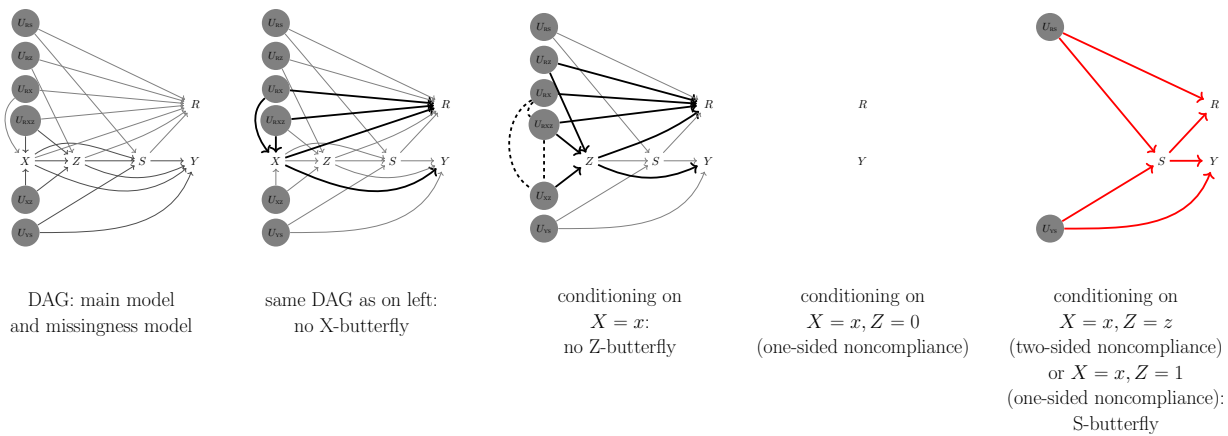}}
\end{figure}

Section~\ref{subsec:general-causal} mentions this case and stated that it does not present any additional conditions on the causal structure for MAR to hold. This is shown in Figure~\ref{fig:third-main}. Specifically in this case there butterfly structure centering $S$ exists, and like in the simple main model, there are no butterfly structure centering $X$ or $Z$.

\subsection{The general main model case}
\label{asubsec:general-main}

See Figure~\ref{fig:general-main-full}, which shows the X-butterfly structure in addition to the S-butterfly structure, and how both structures result in conditional dependence between $R$ and $Y$, through a series of graphs with incremental conditioning (on $X$, then additionally on $Z$, then additionally on $S$).

\begin{figure}
\caption{The general main model combined with the simple missingness model minus type 1 and type 2 paths, shown in unconditional and conditional graphs. Blue (red) color highlights butterfly structure centering $X$ ($S$) and $R$-$Y$ dependence when conditioning on $X$ ($S$).}
\label{fig:general-main-full}
\resizebox{\textwidth}{!}{\input{figures/v2tkz/figS5.tkz}}
\end{figure}

\section{Appendix to Section~\ref{sec:recovery}}
\label{asec:recovery}

\subsection{Connecting the model in Figure~\ref{fig:id-assumptions}A to exclusion restriction}
\label{asubsec:exclusion}

See Figure~\ref{fig:exclusion-restriction}. We start with the principal stratification graph based on the DAG. First, condition on $X$ to arrive at the first conditional graph. Second, we condition additionally on $S(1)=S(0)$. This is, generally, conditioning on a small range rather than a specific value of variable $C$, so we keep the node for $C$ and the graph stays in the same form. Here, however, we can use the additional information that if $S(1)=S(0)$ then $S(1),S(0)$ perfectly determine $S$. This severs the link between $Z$ and $S$, and gives us the second conditional graph, which shows that $C\independent Y\mid X,S(1)=S(0)$ (exclusion restriction).

\begin{figure}[h]
    \centering
    \caption{Connecting the DAG in Figure~\ref{fig:id-assumptions}A to exclusion restriction, $Z\independent Y\mid X,S(1)=S(0)$}
\label{fig:exclusion-restriction}
\resizebox{.9\textwidth}{!}{\input{figures/v2tkz/figS6.tkz}}
\end{figure}

\subsection{Connecting the two main models in Figure~\ref{fig:id-assumptions}B to principal ignorability}
\label{asubsec:principal-ignorability}

See Figure~\ref{fig:principal-ignorability}, which starts with the two principal stratification graphs based on the two DAGs of interest. Then conditioning on $X,V$ obtains the conditional graph on the right, which reveals that if one additionally conditions on $Z,S$ then $C$ and $Y$ are independent. That is, principal ignorability holds.

\begin{figure}[h]
    \centering
    \caption{Connecting the two main model DAGs in Figure~\ref{fig:id-assumptions}B to principal ignorability, $C\independent Y\mid X,V,Z,S$}
\label{fig:principal-ignorability}
\resizebox{.9\textwidth}{!}{\input{figures/v2tkz/figS7.tkz}}
\end{figure}

\subsection{Effect identification -- instrumental variable approach}
\label{asubsec:iv-identification}

Here we continue with step 3 of PCE identification, where we left of at the end of Appendix~\ref{asubsec:prelim-aside}. Recall that Step 1 obtained
\begin{align*}
    \E[Y(1)-Y(0)\mid C]=\E_{X\mid C}\{\E[Y\mid X,Z=1,C]-\E[Y\mid X,Z=0,C]\},\tag{\ref{eq:unconfoundedness}}
\end{align*}
and Step 2 obtained, for the one-sided noncompliance setting,
\begin{alignat*}{3}
    &\P(X\mid\text{complier})&&=\P(X)\frac{\P(S=1\mid X,Z=1)}{\E[\P(S=1\mid X,Z=1)]},&&
    \\
    &\P(X\mid\text{noncomplier})&&=\P(X)\frac{\P(S=0\mid X,Z=1)}{\E[\P(S=0\mid X,Z=1)]},&&
    \\
    &\P(\text{complier}\mid X,Z)&&=\P(S=1\mid X,Z=1),&&
    \\
    &\P(\text{noncomplier}\mid X,Z)&&=\P(S=0\mid X,Z=1).&&
\end{alignat*}
and for the two-sided noncompliance setting,
\begin{alignat*}{3}
    &\P(X\mid\text{complier})&&=\P(X)\frac{\P(S=1\mid X,Z=1)-\P(S=1\mid X,Z=0)}{\E[\P(S=1\mid X,Z=1)-\P(S=1\mid X,Z=0)]},&&
    \\
    &\P(X\mid\text{always-taker})&&=\P(X)\frac{\P(S=1\mid X,Z=0)}{\E[\P(S=1\mid X,Z=0)]},&&
    \\
    &\P(X\mid\text{never-taker})&&=\P(X)\frac{\P(S=0\mid X,Z=1)}{\E[\P(S=0\mid X,Z=1)]},&&
    \\
    &\P(\text{complier}\mid X,Z)&&=\P(S=1\mid X,Z=1)-\P(S=1\mid X,Z=0),&&
    \\
    &\P(\text{always-taker}\mid X,Z)&&=\P(S=1\mid X,Z=0),&&
    \\
    &\P(\text{never-taker}\mid X,Z)&&=\P(S=0\mid X,Z=1).&&
\end{alignat*}
The remaining task of step 3 is to identify $\E[Y\mid X,Z=z,C]$ for $z=0,1$ in (\ref{eq:unconfoundedness}), here using assumptions A3 and A4a.

Under A3 (monotonicity), there are no defiers, so those who take the treatment when assigned to control are always-takers and those who do not take the treatment when assigned to treatment are never-takers, i.e.,
\begin{align*}
    \E[Y\mid X,Z=0,\text{always-taker}]
    &=\E[Y\mid X,Z=0,S=1],
    \\
    \E[Y\mid X,Z=1,\text{never-taker}]
    &=\E[Y\mid X,Z=1,S=0].
\end{align*}

Now we use A4b (exclusion restriction), formally $Z\independent Y\mid X,S(1)=S(0)$. First, consider always-takers and never-takers in the two-sided and noncompliers in the one-sided noncompliance setting. Under A4b,
\begin{align*}
    \E[Y\mid X,Z=1,\text{always-taker}]&=\E[Y\mid X,Z=0,\text{always-taker}]=\E[Y\mid X,Z=0,S=1],
    \\
    \E[Y\mid X,Z=0,\text{never-taker}]&=\E[Y\mid X,Z=1,\text{never-taker}]=\E[Y\mid X,Z=1,S=0],
    \\
    \E[Y\mid X,Z=0,\text{noncomplier}]&=\E[Y\mid X,Z=1,\text{noncomplier}]=\E[Y\mid X,Z=1,S=0],
\end{align*}
so AACE = 0 and NACE = 0.

Now consider compliers. In the two-sided noncompliance setting
\begin{align*}
    &\E[Y\mid X,Z=1]
    =\underbrace{\P(\text{complier}\mid X,Z=1)}_{\P(S=1\mid X,Z=1)-\P(S=1\mid X,Z=0)}\E[Y\mid X,Z=1,\text{complier}]+
    \\
    &~\underbrace{\P(\text{always-taker}\mid X,Z=1)}_{\P(S=1\mid X,Z=0)}\underbrace{\E[Y\mid X,Z=1,\text{always-taker}]}_{\E[Y\mid X,Z=0,S=1)]}+
    \underbrace{\P(\text{never-taker}\mid X,Z=1)}_{\P(S=0\mid X,Z=1)}\underbrace{\E[Y\mid X,Z=1,\text{never-taker}]}_{\E[Y\mid X,Z=1,S=0]}
    \\
    &\Longrightarrow\E[Y\mid X,Z=1,\text{complier}]=
    \\
    &~~~~~~~~~~\frac{\E[Y\mid X,Z=1]-\P(S=1\mid X,Z=0)\E[Y\mid X,Z=0,S=1)-\P(S=0\mid X,Z=1)\E[Y\mid Z=1,S=0]]}{\P(S=1\mid X,Z=1)-\P(S=1\mid X,Z=0)}.
\end{align*}
Similarly,
\begin{align*}
    &\E[Y\mid X,Z=0,\text{complier}]=
    \\
    &~~~~~~~~~~\frac{\E[Y\mid X,Z=0]-\P(S=1\mid X,Z=0)\E[Y\mid X,Z=0,S=1)-\P(S=0\mid X,Z=1)\E[Y\mid Z=1,S=0]]}{\P(S=1\mid X,Z=1)-\P(S=1\mid X,Z=0)}.
\end{align*}
Taking the difference, we have
\begin{align*}
    \E[Y\mid X,Z=1,\text{complier}]-\E[Y\mid X,Z=0,\text{complier}]=\frac{\E[Y\mid X,Z=1]-\E[Y\mid X,Z=0]}{\P(S=1\mid X,Z=1)-\P(S=1\mid X,Z=0)}.
\end{align*}
Plugging this and the result for $\P(X\mid\text{complier})$ into (\ref{eq:unconfoundedness}), we have
\begin{align*}
    \text{CACE}
    &=\E\left\{\frac{\E[Y\mid X,Z=1]-\E[Y\mid X,Z=0]}{\P(S=1\mid X,Z=1)-\P(S=1\mid X,Z=0)}\frac{\P(S=1\mid X,Z=1)-\P(S=1\mid X,Z=0)}{\E[\P(S=1\mid X,Z=1)-\P(S=1\mid X,Z=0)]}\right\}
    \\
    &=\frac{\E\{\E[Y\mid X,Z=1]-\E[Y\mid X,Z=0]\}}{\E[\P(S=1\mid X,Z=1)-\P(S=1\mid X,Z=0)]}.
\end{align*}
In the two-sided noncompliance setting, the reasoning is similar but simpler, and the result is
\begin{align*}
    \text{CACE}=\frac{\E\{\E[Y\mid X,Z=1]-\E[Y\mid X,Z=0]\}}{\E[\P(S=1\mid X,Z=1)]}.
\end{align*}
The job is done!

A comment: This derivation takes longer than the usual derivation which notes that because the other effect(s) is (are) zero, the CACE is a multiplier of the ATE. The purpose of this derivation here, though, is to clearly show that the role of assumption A4 is to help identify the $\E[Y\mid X,Z=z,C]$ component in formula (\ref{eq:unconfoundedness}). Making explicit the conditional outcome means for compliers also has the advantage of helping to check if the assumption is to be trusted. When it does not hold, it may (but also may not) conflict with the observed data distribution, for example, implying $\E[Y\mid X,Z=z,C]$ values that are outside the outcome range.

% As mentioned in the paper, this is included just for completeness.

% Under A0-A2, the ATE ($\E[Y(1)-Y(0)]$) is identified as
% \begin{align*}
%     \text{ATE}=\E\{\E[Y\mid X,Z=1]-\E[Y\mid X,Z=0]\}.
% \end{align*}
% Note that the ATE is a weighted average of the PCEs, where the weights are the sizes of the principal strata. Under exclusion restriction (A3a), AACE and NACE in the two-sided noncompliance setting and NACE in the one-sided noncompliance setting are all zero. Therefore $\text{CACE}=\text{ATE}/\P(C=c)$. The denominator $\P(C=c)$ is identified by $\E[\P(S=1\mid X,Z=1)-\P(S=1\mid X,Z=0)]$ in the two-sided noncompliance setting and by $\E[\P(S=1\mid X,Z=1)]$ in the one-sided noncompliance setting (see Appendix section~\ref{asubsec:prelim-aside}). This obtains: under A0, A1, A2, A3a,
% \begin{align*}
%     \text{CACE}=
%     \begin{cases}
%         \displaystyle\frac{\E\{\E[Y\mid X,Z=1]-\E[Y\mid X,Z=0]\}}{\E[\P(S=1\mid X,Z=1)-\P(S=1\mid X,Z=0)]} & \text{if two-sided noncompliance}
%     \\
%     \\
%     \displaystyle\frac{\E\{\E[Y\mid X,Z=1]-\E[Y\mid X,Z=0]\}}{\E[\P(S=1\mid X,Z=1)]} & \text{if one-sided noncompliance}
%     \end{cases}.
% \end{align*}

\subsection{Effect identification -- principal ignorability approach}
\label{asubsec:pi-identification}

For simplicity of presentation, we start with the simpler version of principal ignorability ($C\independent Y\mid X,Z,S$) here, and will focus on the two-sided noncompliance setting. Reasoning for the one-sided noncompliance setting is similar and simpler. We already have all the results of steps 1 and 2 that have been copied to Appendix~\ref{asubsec:iv-identification} above. We have also argued in that section that under monotonicity (A3),
\begin{align*}
    \E[Y\mid X,Z=0,\text{always-taker}]
    &=\E[Y\mid X,Z=0,S=1],
    \\
    \E[Y\mid X,Z=1,\text{never-taker}]
    &=\E[Y\mid X,Z=1,S=0].
\end{align*}
Now, under principal ignorability, we have
\begin{align*}
    \E[Y\mid X,Z=1,\text{always-taker}]=\E[Y\mid X,Z=1,\text{complier}]&=\E[Y\mid X,Z=1,S=1],
    \\
    \E[Y\mid X,Z=0,\text{never-taker}]=\E[Y\mid X,Z=0,\text{complier}]&=\E[Y\mid X,Z=0,S=0].
\end{align*}
Therefore
\begin{align*}
    \E[Y\mid X,Z=1,C]-\E[Y\mid X,Z=0,C]
    =
    \begin{cases}
        \E[Y\mid X,Z=1,S=1]-\E[Y\mid X,Z=0,S=0] & \text{if}~C=\text{complier}
        \\
        \E[Y\mid X,Z=1,S=1]-\E[Y\mid X,Z=0,S=1] & \text{if}~C=\text{always-taker}
        \\
        \E[Y\mid X,Z=1,S=0]-\E[Y\mid X,Z=0,S=0] & \text{if}~C=\text{never-taker}
    \end{cases}.
\end{align*}
Combining this with the results for the stratum-specific covariate distributions, we have
\begin{align*}
    \text{CACE}&=\E\left(\big\{\E[Y\mid X,Z=1,S=1]-\E[Y\mid X,Z=0,S=0]\big\}\frac{\P(S=1\mid X,Z=1)-\P(S=1\mid X,Z=0)}{\E[\P(S=1\mid X,Z=1)-\P(S=1\mid X,Z=0)]}\right),
    \\
    \text{AACE}&=\E\left(\big\{\E[Y\mid X,Z=1,S=1]-\E[Y\mid X,Z=0,S=1]\big\}\frac{\P(S=1\mid X,Z=0)}{\E[\P(S=1\mid X,Z=0)]}\right),
    \\
    \text{NACE}&=\E\left(\big\{\E[Y\mid X,Z=1,S=0]-\E[Y\mid X,Z=0,S=0]\big\}\frac{\P(S=0\mid X,Z=1)}{\E[\P(S=0\mid X,Z=1)]}\right).
\end{align*}

Under the general principal ignorability assumption, there are two ways to derive identification results. One is to keep all the results from steps 1 and 2, and use the assumption to identify $\E[Y\mid X,Z=z,C]$; this is quite complicated. The simpler way is to notice that the combination of $(X,V)$ also satisfies treatment assignment ignorability, formally $Z\independent(S(1),S(0),Y(1),Y(0))\mid X,V$, and treatment assignment positivity, formally $0<\P(Z=1\mid X,V)<1$. This allows using the results above but replacing $X$ with $X,V$.

\end{document}